\documentclass[twocolumn,amsmath,amssymb]{revtex4}

\usepackage{graphicx}
\usepackage{dcolumn}
\usepackage{bm}

\begin{document}

\title{The effect of short ray trajectories on the scattering statistics of wave chaotic systems}

\author{J. A. Hart}
\author{T. M. Antonsen, Jr}
\author{E. Ott}

\affiliation{
University of Maryland\\
College Park, MD 20740 }

\date{\today}

\begin{abstract}
In many situations, the statistical properties of wave systems with chaotic classical limits are well-described by random matrix theory.  However, applications of random matrix theory to scattering problems require introduction of system specific information into the statistical model, such as the introduction of the average scattering matrix in the Poisson kernel.  Here it is shown that the average impedance matrix, which also characterizes the system-specific properties, can be expressed in terms of classical trajectories that travel between ports and thus can be calculated semiclassically.  Theoretical results are compared with numerical solutions for a model wave-chaotic system.
\end{abstract}
\maketitle




\section{Introduction}

Wave systems appear in many different branches of physics, such as quantum mechanics, classical electromagnetism and acoustics.  However, solving the wave equations in general can be quite difficult, particularly in the short wavelength limit for systems which have chaotic dynamics in the classical limit \cite{gutzwiller_book}.  Furthermore, even if exact solutions were feasible, there may be uncertainties in the locations of boundaries or in parameters specifying the system.  Thus, rather than solving such systems exactly, it has often been convenient to create statistical models which reproduce the generic properties of such systems without the need to accurately model the details \cite{haake_book}.  One successful statistical approach, known as random matrix theory, is to replace the exact wave-mechanical operators, such as the Hamiltonian or scattering matrix, with matrices whose elements are assumed to be random.  Although such formulations cannot predict any particular wave system's properties exactly, they can predict the distribution of properties in an ensemble of related wave-chaotic systems.  Random matrix theory also predicts the statistical properties of a \emph{single} wave-chaotic system evaluated at different frequencies.  The random matrix technique applies to a wide range of systems and has been well studied both theoretically and experimentally.  See Refs.~\cite{Alhassid_review,Guhr_WeidenMueller_MuellerGroeling_review,Quantum_graphs_review,Beenakker_review_RMT} for reviews of the theory, history and applications of random matrix theory.

In this paper, we use random matrix theory to model the scattering behavior of an ensemble of wave-chaotic systems coupled to the outside world through $M$ discrete scattering channels.  Such scattering systems have been studied extensively, with most work focusing on the $M\times M$ scattering matrix $\boldsymbol{S}$, either by using a random Hamiltonian for the closed system and deriving the resulting scattering matrix using assumptions for the coupling between the wave system and the scattering channels \cite{Brouwer_Lorentzian} or by replacing the scattering matrix with a random matrix directly \cite{Dyson_original,Doron_Smilansky_Poisson,Poisson_Kernel_Original,Poisson_including_internal,Poisson_including_internal_2}.  These two approaches are complementary and for some ensembles have been explicitly shown to be equivalent \cite{Brouwer_Lorentzian}.

We consider ensembles of systems whose distribution of scattering matrices are well-described by the so-called Poisson kernel \cite{Dyson_original,Poisson_Kernel_Original,Doron_Smilansky_Poisson}.  The Poisson kernel characterizes the probability density for observing a particular scattering matrix $\boldsymbol{S}$ in terms of the average scattering matrix $\boldsymbol{\bar{S}}$, which is also called the `optical scattering matrix'.  It represents contributions to the scattering matrix from elements of the system which are not random.  For instance, if the scattering channels are not perfectly coupled to the wave system, some fraction of the energy in the incident waves will simply bounce off the interface between the channel and the scatterer without experiencing the chaotic aspects of the scatterer, thus strongly constraining $\boldsymbol{\bar{S}}$.  This is known as the prompt reflection \footnote{In nuclear scattering, the scattering dynamics naturally splits into two different timescales, with reflections due to poor coupling and so-called ``direct processes'' occurring very quickly compared to the slow ``equilibrated response.''  In these cases, the direct processes are usually included in the term ``prompt reflection.''  In microwave billiards, the distinction between direct processes and equilibrated response is much less clear because the relevant time-scales are not nearly as well separated.  However, in microwave billiards, the size of the ports used to couple into the system are typically much smaller than any of the other distances in the system and therefore respond much more quickly to incident waves.  Thus we designate only the reflections due to the dynamics within the ports as the prompt response.}.  In addition, rays within the scattering region which connect the scattering channels without ergodically exploring the chaotic dynamics also affect $\boldsymbol{\bar{S}}$ \cite{Poisson_including_internal,Bulgakov_Gopar_Mello_Rotter}.

Because $\boldsymbol{\bar{S}}$ is the only parameter in the Poisson kernel, methods of finding it for a specific system are of interest.  Although $\boldsymbol{\bar{S}}$ can be extracted quite simply from experimental data, \emph{predicting} it from first principles is quite difficult in general, although it has been done for for some specific systems such as quantum graphs \cite{Kottos_Smilansky_2003}. In most wave systems, however, it depends in a complicated way on the interactions between the scattering channels, the wave system, and any significant classical trajectories.  To address this problem, we find it convenient to transform from the scattering matrix $\boldsymbol{S}$ to the $M\times M$ impedance matrix $\boldsymbol{Z}$ \cite{henrythesis,Henry_paper_one_port,Henry_paper_many_port},
\begin{equation}\label{eq:impedance_from_scattering}
    \boldsymbol{Z}=\boldsymbol{Z}_{0}^{1/2}(\boldsymbol{1}+\boldsymbol{S})(\boldsymbol{1}-\boldsymbol{S})^{-1}\boldsymbol{Z}_{0}^{1/2}.
\end{equation}
where $\boldsymbol{Z}_0$ is an $M\times M$ diagonal matrix whose $i$th diagonal element is determined by the detailed properties of the $i$th scattering channel as described below.  We will show that the average impedance matrix (to be defined) can be expressed directly in terms of classical ray trajectories.

Impedance is a meaningful concept for all scattering wave systems.  In linear electromagnetic systems, it is defined via the phasor generalization of Ohm's law as
\begin{equation}\label{eq:Ohms_law}
    \hat{V}=\boldsymbol{Z}\hat{I},
\end{equation}
where the $M$-dimensional vector $\boldsymbol{\hat{V}}$ represents the voltage differences across the attached transmission lines (the systems port) and the $M$-dimensional vector $\boldsymbol{\hat{I}}$ denotes the currents flowing through the transmission lines.  The concept of impedance can be generalized to cases where the cavity is excited through apertures connected to waveguides that may support several propagating modes.  In acoustics, the impedance is the ratio of the sound pressure to the fluid velocity.  A quantum-mechanical quantity corresponding to impedance is the reaction matrix, which is often denoted in the literature as $i\boldsymbol{K}$ \cite{Alhassid_review}.

The diagonal elements of $\boldsymbol{Z}_{0}$ are the characteristic impedances of the scattering channels.  In electromagnetic systems with transmission lines for scattering channels, the characteristic impedances are the ratio between the voltage difference across the transmission line and the current through the transmission line for a monochromatic wave propagating a single direction through the transmission line.  Other wave systems have analogous definitions for $\boldsymbol{Z}_{0}$ determined by the details of the scattering channels.  In what follows, we use terminology appropriate in the context of an electromagnetic cavity connected to the outside world via transmission line channels.

With the transformation to impedance, we find that we can define an ``average'' impedance matrix $\boldsymbol{Z}_{avg}$ which is related to $\boldsymbol{\bar{S}}$ via the transformation,
\begin{equation}\label{eq:average_impedance_from_scattering}
    \boldsymbol{Z}_{avg}=\boldsymbol{Z}_{0}^{1/2}(\boldsymbol{1}+\boldsymbol{\bar{S}})(\boldsymbol{1}-\boldsymbol{\bar{S}})^{-1}\boldsymbol{Z}_{0}^{1/2}.
\end{equation}
In contrast to $\boldsymbol{\bar{S}}$, we find that $\boldsymbol{Z}_{avg}$ can be evaluated directly in the semiclassical limit as a sum over contributions from the prompt reflection and short classical trajectories.  Thus, through \eqref{eq:average_impedance_from_scattering} this gives a method for approximating $\boldsymbol{\bar{S}}$ in the semiclassical limit.

In this paper, we present our approach to calculating $\boldsymbol{Z}_{avg}$.  In Sec.~\ref{sec:Overview}, we provide an overview of our theory for lossless systems and describe the most important results of our investigations.  In Sec.~\ref{sec:semiclassical-theory}, we find expressions for the impedance of a specific quasi-2D lossless microwave cavity as explicit functions of the boundaries and port positions, creating a framework in which we can keep some cavity properties fixed and let others change.  We then apply the semiclassical approximation to our exact formulations.  In Sec.~\ref{subsec:short-orbit-approx}, we use the semiclassical approximation to derive expressions for the impedance induced by objects near the port in terms of classical short orbits between the ports and the internal scatterers.  In Sec.~\ref{subsec:finite_matrix}, we use the semiclassical approximation to convert our exact solution with integral operators into a finite-dimensional matrix equation with an internal scattering matrix $\boldsymbol{T}$.  In Sec.~\ref{sec:statistics}, we assume that the matrix $\boldsymbol{T}$ is distributed according to the Poisson kernel with the average $\boldsymbol{T}$ given by the results of Sec.~\ref{sec:semiclassical-theory} and use a result by Brouwer \cite{Brouwer_Lorentzian} to find the corresponding distribution for the impedance.  In Sec.~\ref{sec:numerical-testing}, we demonstrate that in the lossless case, our theory agrees with numerical simulations of our system.  In Sec.~\ref{sec:adding_loss}, we extend our theory to lossy cavities and briefly refer to experimental results in lossy systems which will be published separately.

\section{Overview of Lossless Theory}
\label{sec:Overview}

The results presented in this paper are an extension of our previously developed random coupling model \cite{Henry_paper_one_port,Henry_paper_many_port,henrythesis}.  For simplicity, we consider only systems and frequencies in which the scattering channels have a single propagating mode and in which the ports which couple the scattering channels to the cavity are separated from each other by much more than a wavelength.  We found previously with these assumptions that by replacing the resonant frequencies of our closed chaotic cavity with a set of resonant frequencies appropriate to a random matrix drawn from the Gaussian orthogonal ensemble and modeling the eigenfunctions using the random-plane wave hypothesis, the impedance of our lossless wave-chaotic systems could be described by \cite{Henry_paper_many_port}
\begin{equation}\label{eq:chaotic_impedance_statistics}
    \boldsymbol{Z}=i\boldsymbol{X}_{R}+i\sqrt{\boldsymbol{R}_{R}}\boldsymbol{\xi}_{0}\sqrt{\boldsymbol{R}_{R}},
\end{equation}
where $\boldsymbol{R}_{R}$, the radiation resistance, and $\boldsymbol{X}_{R}$, the radiation reactance, are the real and imaginary parts of the $M\times M$ diagonal radiation impedance matrix $\boldsymbol{Z}_{R}$, which represents the impedance the scattering system would have if all the energy which successfully coupled into the system was absorbed rather than allowed to couple back out. The matrix $\boldsymbol{\xi}_{0}$ is an element of the appropriate $M\times M$ Lorentzian ensemble introduced by Brouwer \cite{Brouwer_Lorentzian} with width 1 and median 0, which in the single-port case simplifies to a Lorentzian random variable with width 1 and median 0.  We denote $\boldsymbol{\xi}_{0}$ the normalized impedance and have previously studied its properties in chaotic systems \cite{henrythesis,Henry_paper_one_port,Henry_paper_many_port,Experimental_tests_our_work}.

Equation~\eqref{eq:chaotic_impedance_statistics} is the direct impedance analog of the Poisson kernel distribution for $\boldsymbol{S}$ in the case that the only contribution to $\boldsymbol{\bar{S}}$ is due to the `prompt reflections' caused by the impedance mismatch between the scattering channels and the wave system \cite{henrythesis}.  From experimental measurements and simulation results performed using the commercial off-the shelf program High Frequency Structure Simulator(HFSS), we know that Eq.~\eqref{eq:chaotic_impedance_statistics} describes the impedance statistics of our sample systems only if the impedances are sampled from a very wide frequency range \cite{henrythesis}.  We find, however, that if we consider sample impedances from either narrower frequency ranges or from many slightly different chaotic systems, the distribution of the resulting impedances is still well-described by a Lorentzian distribution, but with a different median and width than that predicted by Eq.~\eqref{eq:chaotic_impedance_statistics}.

\begin{figure}
\begin{center}
\includegraphics[width=3in]{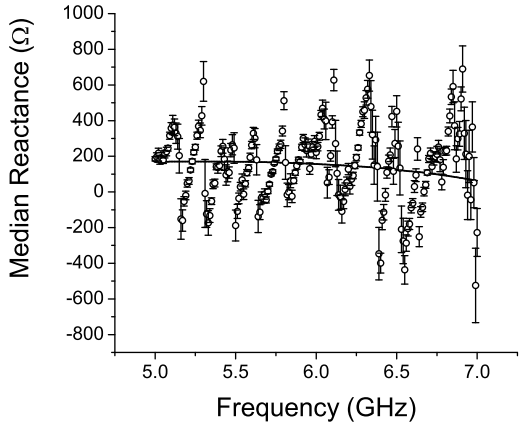}
\end{center}
\renewcommand{\baselinestretch}{1}
\small\normalsize
\begin{quote}
\caption{\label{fig:median-deviation}
A comparison between the port radiation reactance as measured by HFSS (solid line) and the ensemble median of the HFSS simulated impedances (circles).  The random coupling model predicts that if the ensemble is sufficiently random, the ensemble median should equal the radiation reactance.  The error bars were estimated by assuming that the ensemble impedance is a Lorentzian random variable (justified by statistical examination of the ensemble data) and finding the uncertainty in the median, given the numerically found width.  The differences between these two curves are caused by short orbits within the cavity which exist in many realizations of the ensemble.}
\end{quote}
\end{figure}

\begin{figure}
\begin{center}
\includegraphics[width=3in]{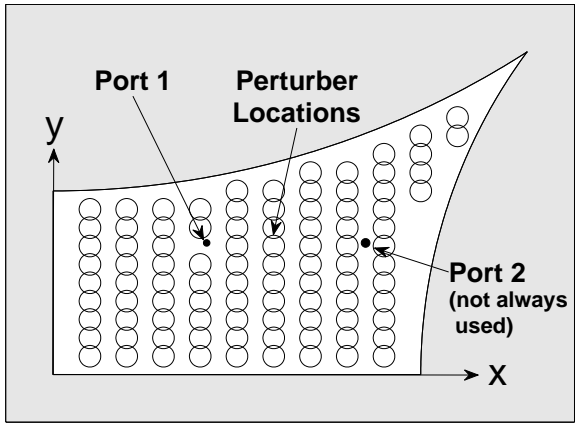}
\end{center}
\renewcommand{\baselinestretch}{1}
\begin{quote}
\caption{\label{fig:ensemble-picture}
This plot displays a 2-D view of our simulated microwave cavity and the perturber positions used to produce the ensemble displayed in Fig.~\ref{fig:median-deviation}.  The outer walls are fixed in all realizations, while every realization has the perturber at a different location.  The microwaves are fed into the cavity through the ports, which are coaxial cables inserted through the top of the cavity.  The dimension $h$ of the cavity in the z-direction (out of the page) are much smaller than the wavelengths used to excite the cavity and therefore results in effectively two-dimensional waves in the x-y plane.}
\end{quote}
\end{figure}

These deviations are illustrated in Fig.~\ref{fig:median-deviation}, which shows the median calculated impedance for the quasi-two dimensional cavity illustrated in Fig.~\ref{fig:ensemble-picture} (its dimensions are given in Sec.~\ref{sec:numerical-testing}).  This cavity was the basis for our previous numerical and experimental research.  It is a simulated electromagnetic cavity filled with a uniform lossless dielectric and is coupled to the outside world through coaxial cables (the ports) inserted into holes on the top of the cavity.  For the data in Fig.~\ref{fig:median-deviation}, only port 1 is present.  On the walls we impose perfect-conductor boundary conditions.  Because our cavity has a uniform height $h$ in the z-direction which is much smaller than the wavelength of the incident microwaves, Maxwell's equations become effectively two-dimensional with the electric and magnetic fields uniform in the z-direction \cite{Original_microwave_resonator,Henry_paper_one_port}.  This system is an example of a wave billiard, meaning that the rays within the cavity follow straight lines except for specular reflection at the walls.  To produce the simulation data shown in Fig.~\ref{fig:median-deviation}, we generated 95 different realizations of related systems by adding a small mobile perturber to our baseline system and moving it to 95 different, widely-spaced locations (see Fig.~\ref{fig:ensemble-picture}).  We then find that, at each frequency, the distribution of impedances is Lorentzian, but with a median and width which are different from $\boldsymbol{X}_{R}$ and $\boldsymbol{R}_{R}$.  (For this example, $\boldsymbol{X}_{R}$ and $\boldsymbol{R}_{R}$ are scalars.)  We also find that as a function of frequency, the fitted medians and widths oscillate almost symmetrically around $\boldsymbol{X}_{R}$ and $\boldsymbol{R}_{R}$.  As we will later see, this behavior arises because short classical trajectories within the system alter the distribution of impedances, analogous to the distortion of $\boldsymbol{\bar{S}}$ observed in previous work \cite{Poisson_including_internal,Poisson_including_internal_2,Bulgakov_Gopar_Mello_Rotter}.

In this paper we show that corrections to the radiation impedance matrix due to the direct orbits redefine $\boldsymbol{Z}_{R}$ in an additive way, $\boldsymbol{Z}_{R}\rightarrow \boldsymbol{Z}_{avg}= \boldsymbol{Z}_{R}+\left[\textrm{direct orbit terms}\right]$.  More specifically, we find that
\begin{equation}\label{eq:Z_avg_introduction}
    \boldsymbol{Z}_{avg}=\boldsymbol{Z}_{R}+\boldsymbol{R}_{R}^{1/2}\boldsymbol{\zeta}\boldsymbol{R}_{R}^{1/2},
\end{equation}
where $\boldsymbol{\zeta}$ is an $M\times M$ dimensionless matrix whose $(m,n)$th element describes the effects of wave propagation from port $m$ to port $n$ and is explicitly defined in Eq.~\eqref{eq:zeta_def}.  The individual elements $\zeta_{m,n}$ are each a sum over all different possible ray paths going from port $m$ to port $n$, with each path having a phase factor proportional to the length of the path and prefactors describing the directivity of the classical trajectories relative to the shape of the ports, the stability of the ray trajectory and the number of reflections from the boundaries.  In addition, in cases where the geometry of the cavity is varied to create an ensemble (for example by moving a perturber throughout the cavity) there is a factor that accounts for the fraction of realizations in which that particular path will contribute (i.e., not be blocked by the perturber).  We note that as the frequency window within which sample impedances are generated gets wider and/or the ensemble changes more drastically between realizations, the value of $\boldsymbol{\zeta}$ needed to normalize the data goes to $0$ and we get our original random coupling model back as a limit.

The substitution of $\boldsymbol{Z}_{avg}$ for $\boldsymbol{Z}_{R}$ can be understood to be a generalization of the radiation impedance of the ports to include the effects of features of the cavity that are distant from the ports but that do not vary from one member of the ensemble to another.  Consider an ensemble of lossless microwave cavities whose generic properties (such as volume and circumference) are fixed but whose shapes are random and independent \emph{except} that the ports are always placed in the same positions relative each other and except that some segments of the wall are also fixed.  The impedance of each configuration will reflect the interaction between the ports and both the fixed and varying segments at the wall.  The interactions between the ports and fixed wall segments will be the same for each member of the ensemble and will contribute to the average impedance while the interactions between the ports and the varying segments will vary from member to member and contribute to statistical deviations from the average.  As our subsequent analysis will show, the ports and fixed wall segments can together be considered to be a single super-port which has a radiation impedance of $\boldsymbol{Z}_{avg}$, thus justifying replacing $\boldsymbol{Z}_{R}$ with $\boldsymbol{Z}_{avg}$ in Eq.~\eqref{eq:chaotic_impedance_statistics}.

\section{The Impedance as a Function of Cavity Shape}
\label{sec:semiclassical-theory}

For the electromagnetic system described in Sec.~\ref{sec:Overview}, we previously derived \cite{Henry_paper_many_port} the following inhomogeneous wave equation for the case where the ports are modeled by vertical (z-direction), externally imposed, localized current densities flowing from the bottom to the top plates,
\begin{equation}\label{eq:Helmholtz_equation_local}
    \left(\nabla_{\bot}^{2}+k^2\right)\hat{V}_{T}(\vec{r})=i k h \eta\sum_{p=1}^{M}u_{p}(\vec{r})I_p,
\end{equation}
where $\nabla_{\bot}$ is the 2D Laplacian in the $(x,y)$ plane,$\hat{V}_{T}$ represents the voltage difference between the two plates, $I_p$ represents the total current injected into the cavity through port $p$, $u(\vec{r})$ represents the profile of the current injected onto the top plate at port $p$ and has the property $\int d\vec{r}'\,u(\vec{r}')=1$, $\eta=\sqrt{\mu/\epsilon}$ is the wave impedance of propagation within the medium inside the cavity (i.e.\ $\eta$ is the ratio of the electric field to the magnetic field in an infinite plane wave), and $k=2\pi/\lambda$ is the wave-number of the external driving frequency.  With perfect-conducting boundary conditions, $V_{T}$ must be zero on the cavity boundary.  The model of the port considered here is appropriate to the case in which the port is smaller than a wavelength so that the distribution of current (given by $u(\vec{r}')$) is fixed, independent of frequency of the fields in the cavity $V_{T}(\vec{r})$.  In this way each port is characterized by a single current $I_{p}$, and a corresponding voltage \cite{Henry_paper_many_port},
\begin{equation}\label{eq:voltage_def}
    V_{p}=\int d^2\vec{r}'\,u_{p}(\vec{r}')V_{T}(\vec{r}').
\end{equation}
Definition~\eqref{eq:voltage_def} was selected in Ref.~\cite{Henry_paper_one_port} since it yields $P=(1/2)\textrm{Re}\{V_{p}^{*}I_{p}\}$ for the power flow into the cavity.  The cavity impedance then gives the matrix relation between the port currents $I_{p}$ and port voltages $V_{p}$.

Equation~\eqref{eq:Helmholtz_equation_local} is the driven Helmholtz equation, and although it was derived in the context of quasi-2D electromagnetic cavities and a particular port model, it can be applied to many different types of systems (such as quantum dots or acoustic resonators) simply by relabeling the constants and tweaking the boundary conditions \cite{Original_microwave_resonator}.

Before considering statistics, we first derive an expression for the impedance for individual realizations of the cavity.  Similar to previous work by Georgeot and Prange \cite{first_fredholm_georgeot_prange}, we can convert Eq.~\eqref{eq:Helmholtz_equation_local} into an integral equation.  We do this by introducing the outgoing Green's function $G_0(\vec{r},\vec{r}';k)$ which satisfies
\begin{equation}\label{eq:outgoing_green}
    \left(\nabla_{\bot}^{2}+k^2\right)G_0(\vec{r},\vec{r}';k)=\delta(\vec{r}-\vec{r}').
\end{equation}
We then multiply both sides of Eq.~\eqref{eq:Helmholtz_equation_local} by $G_0(\vec{r},\vec{r}';k)$ and integrate both sides over $\vec{r}'$ obtaining
\begin{eqnarray}\label{eq:integrated_Helmholtz_equation}
    \int_{\mathcal{D}} d^2\vec{r}'G_{0}(\vec{r},\vec{r}')\left(\nabla_{\bot}^{'2}+k^2\right)\hat{V}_{T}(\vec{r}')=&\\ i k h \eta\sum_{p=1}^{M}&I_p \int_{\mathcal{D}} d^2\vec{r}' G_{0}(\vec{r},\vec{r}')u_{p}(\vec{r}'),\nonumber
\end{eqnarray}
where $\mathcal{D}$ denotes the two-dimensional domain within the cavity.  Applying Green's second identity in two dimensions to the left-hand side of Eq.~\eqref{eq:integrated_Helmholtz_equation} and applying the boundary condition on $\hat{V}(\vec{r})$, this becomes
\begin{eqnarray}\label{eq:internal_integral_equation}
    \hat{V}_{T}(\vec{r})&=&-\int_{\partial\mathcal{D}} dq'\,G_{0}(\vec{r},q')\frac{\partial \hat{V}_{T}(q')}{\partial n'}\\&&+i k h \eta\sum_{p=1}^{M}I_p\int_{\mathcal{D}} d^2\vec{r}' G_{0}(\vec{r},\vec{r}')u_{p}(\vec{r}')\nonumber
\end{eqnarray}
where $q'$ represents a position on the boundary $\partial\mathcal{D}$ of the cavity and the integral over $q'$ integrates over the cavity boundary $\partial\mathcal{D}$, and where $\partial/\partial n'$ denotes a derivative in the direction normal to the surface of the cavity at $q'$.

In electromagnetic systems, Eq.~\eqref{eq:internal_integral_equation} has a physical interpretation.  From Maxwell's equations and the perfect conductor boundary conditions, we find that the gradient of the voltage, $\vec{\nabla} \hat{V}_{T}$ is proportional to the surface current in the upper and lower plate of the cavity, with the two currents flowing in opposite directions.  At the edges of the cavity, the surface current flowing in the lower plate travels up the outer wall and into the top plate.  Thus it temporarily travels in the $z$-direction.  We can then interpret Eq.~\eqref{eq:internal_integral_equation} as stating that the electric field inside the cavity is simply the sum over the field radiating from the changing currents in the ports and the field radiating from the changing current in the outer walls.  Therefore, solving Eq.~\eqref{eq:internal_integral_equation} is equivalent to finding the self-consistent current induced in the walls, given the currents in the ports.  We evaluate the normal derivative of Eq.~\eqref{eq:internal_integral_equation} on the surface to get our integral equation,
\begin{eqnarray}\label{eq:external_integral_equation}
    \frac{\partial\hat{V}_{T}(q)}{\partial n}&=&-\int_{\partial\mathcal{D}} dq'\,\frac{\partial G_{0}(q,q')}{\partial n}\frac{\partial \hat{V}_{T}(q')}{\partial n'}\\&&+i k h \eta\sum_{p=1}^{M}I_p\int_{\mathcal{D}} d^2\vec{r}' \frac{\partial G_{0}(\vec{r},\vec{r}')}{\partial n}u_{p}(\vec{r}').\nonumber
\end{eqnarray}
We henceforth drop the subscripts $\partial\mathcal{D}$ and $\mathcal{D}$ on the integral symbols. As long as the function defining the boundary of the cavity is well-behaved, Eq.~\eqref{eq:external_integral_equation} is a Fredholm integral equation of the second type and can be solved via the established Fredholm theory \cite{integral_equations}.

To simply and clarify our results, we follow Prange, Fishman and Georgeot \cite{first_fredholm_georgeot_prange,fredholm_method_for_scars}, and define the operators
\begin{eqnarray}
  \boldsymbol{K}\phi(q) &=& -\int dq'\,\frac{\partial G_{0}(q,q')}{\partial n}\phi(q'), \nonumber\\
  \boldsymbol{V_{+}}u(\vec{r}) &=& \frac{1}{\sqrt{k}}\int d^2\vec{r}'\,\frac{\partial G_{0}(q,\vec{r}')}{\partial n}u(\vec{r}'), \label{eq:operator_defs}\\
  \boldsymbol{V_{-}}\phi(q) &=&  \sqrt{k}\int dq'\,G_{0}(r,q')\phi(q'),\nonumber\\
  \boldsymbol{G_{0}}u(\vec{r}) &=&  \int d^2\vec{r}'\,G_{0}(\vec{r},\vec{r}')u(\vec{r}').\nonumber
\end{eqnarray}
In these operators, $q$ and $q'$ are real scalars denoting points on the cavity boundary $\partial\mathcal{D}$.  They represent distance along the boundary of the cavity as measured relative to some arbitrary starting position.  The vectors $\vec{r}$ and $\vec{r}'$ represent positions within the cavity (i.e.\ within $\mathcal{D}$).  Every operator integrates over a primed variable and maps it onto the space represented by the unprimed variable.

\begin{figure}
\begin{center}
\includegraphics[width=3in]{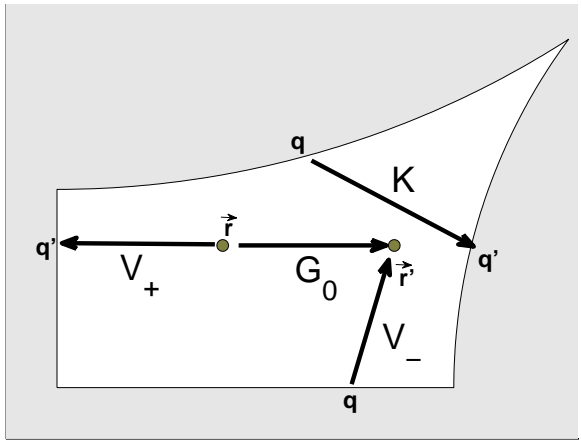}
\end{center}
\renewcommand{\baselinestretch}{1}
\begin{quote}
\caption{\label{fig:propagation_operator_schematic}
A schematic of the different operators defined in Eq.~\eqref{eq:operator_defs}.  Each operator takes a current at the source position and finds the resulting current ($\boldsymbol{K}$ and $\boldsymbol{V}_{+}$) or voltage ($\boldsymbol{G}_0$ and $\boldsymbol{V}_{-}$) induced at the endpoint.}
\end{quote}
\end{figure}

These operators all have physical meanings in electromagnetism.  $G_{0}$  is the two-dimensional, outgoing Green's function in empty space; it finds the voltage at some position $\vec{r}$ caused by a delta-function current distribution at point $\vec{r}'$.  The operator $\boldsymbol{V_{+}}$ finds the current induced in the wall by a delta-function current in the volume.  The operator $\boldsymbol{K}$ represents the current induced in one part of the wall by the current in another part of the wall.  The operator $\boldsymbol{V_{-}}$, on the other hand, gives the voltage inside the volume which results from the currents in the walls.  A schematic of the effects of these operators is shown in Fig.~\ref{fig:propagation_operator_schematic}.

Using this operator notation and solving Eq.~\eqref{eq:external_integral_equation}, we convert Eq.~\eqref{eq:internal_integral_equation} into
\begin{equation}\label{eq:internal_operator_solution}
    \hat{V}_{T}(\vec{r})=i k h\eta\sum_{p=1}^{M}I_p\left[\boldsymbol{V_{-}}\left( \boldsymbol{1}-\boldsymbol{K} \right)^{-1}\boldsymbol{V_{+}}+\boldsymbol{G_{0}}\right]u_{p}(\vec{r}).
\end{equation}
In Fredholm theory, the operator $\left( \boldsymbol{1}-\boldsymbol{K} \right)^{-1}$ is well-defined and can be defined as a ratio of two convergent sums
\begin{equation}\label{eq:fredholm_inversion}
    \left( \boldsymbol{1}-\boldsymbol{K} \right)^{-1}=\frac{\sum_{n=0}^{\infty} \sum_{r=0}^{n}\boldsymbol{K}^{n-r}d_{r}}{\sum_{n=0}^{\infty} d_{n}},
\end{equation}
where $d_{n}$ is an $n$th order polynomial in the traces of $\boldsymbol{K}^{m}$, $m\leq n$.  For more details on constructing $d_{n}$, see the references \cite{first_fredholm_georgeot_prange,fredholm_method_for_scars}, where it is denoted $D_{n}$.

Using the definition of port voltage, Eq.~\eqref{eq:voltage_def}, we get the impedance between ports $n$ and $m$
\begin{equation}\label{eq:first_cavity_impedance}
    Z_{n,m}=i k h \eta\int d^2\vec{r}\,u_{n}(\vec{r})\left[\boldsymbol{V_{-}}\left( \boldsymbol{1}-\boldsymbol{K} \right)^{-1}\boldsymbol{V_{+}}+\boldsymbol{G_{0}}\right]u_{m}(\vec{r}).
\end{equation}
The second term in the integral on the right-hand side of Eq.~\eqref{eq:first_cavity_impedance} represents the impedance the system would have if the walls were moved to infinity and outgoing boundary conditions were imposed but impedance due to direct orbits between the ports were still included.  Therefore, we define an $M\times M$ matrix $\boldsymbol{\tilde{Z}_{R}}$ which has the elements
\begin{equation}\label{eq:radiation_impedance_def}
    \tilde{Z}_{R,n,m}=i k h \eta\int d^2\vec{r}\,u_{n}(\vec{r})\boldsymbol{G_{0}}u_{m}(\vec{r}).
\end{equation}
The diagonal elements of $\boldsymbol{\tilde{Z}}_{R}$ are equal to the diagonal elements of the radiation impedance $\boldsymbol{Z}_{R}$ from Eq.~\eqref{eq:chaotic_impedance_statistics} \cite{Henry_paper_one_port} and the off-diagonal elements represent contributions to the impedance from direct orbits between the ports.  Because the distance between the ports is large compared to a wavelength, we can treat the off-diagonal terms semiclassically.  The diagonal terms $\tilde{Z}_{R,n,n}$ depend on near-field interactions within the port and thus are sensitive to the detailed properties of the port. As in Eq.~\eqref{eq:chaotic_impedance_statistics}, rather than attempting to solve the for the diagonal elements of $\boldsymbol{\tilde{Z}}_{R}$, we treat them as inputs to the theory.  This has the advantage of freeing us from a detailed port model; we expect our model to be accurate even when the port behavior is not modeled by Eq.~\eqref{eq:Helmholtz_equation_local}.

In addition, we define the corresponding radiation resistance and radiation reactance matrices as
\begin{eqnarray}\label{eq:radiation_reactance_resistance}
    \boldsymbol{\tilde{R}}_{R}=\frac{1}{2}\left(\boldsymbol{\tilde{Z}}_{R}+\boldsymbol{\tilde{Z}}_{R}^{\dag}\right),\\
    \boldsymbol{\tilde{X}}_{R}=-\frac{i}{2}\left(\boldsymbol{\tilde{Z}}_{R}-\boldsymbol{\tilde{Z}}_{R}^{\dag}\right),
\end{eqnarray}
where $\dag$ denotes the conjugate transpose.  In the case of the Helmholtz equation, or any system in which time-reversal symmetry is present, $\boldsymbol{\tilde{Z}}_{R}$ is symmetric and so $\boldsymbol{\tilde{R}}_{R}$ and $\boldsymbol{\tilde{X}}_{R}$ are simply the real and imaginary components of $\boldsymbol{\tilde{Z}}_{R}$.

Equation~\eqref{eq:first_cavity_impedance} is an exact solution to Eq.~\eqref{eq:Helmholtz_equation_local} explicitly in terms of the boundaries.  Analytically it is intractable.  In the following two sections, we consider two different, but equivalent, approximations to Eq.~\eqref{eq:first_cavity_impedance}.  It is by equating these two different formalisms that we derive our refined theory.

\subsection{The short-orbit formulation}
\label{subsec:short-orbit-approx}

To get useful theoretical results from Eq.~\eqref{eq:first_cavity_impedance}, we make the assumption that each port $p$ is located near the position $\vec{r}_{0,p}$ and that $u(\vec{r})$ is nonzero only within a small radius around $\vec{r}_{0,p}$.  We assume that the ports are physically separated from each other and from the walls by much more than a wavelength, and that the dimensions of the cavity as a whole are much larger than a wavelength.  With these assumptions, we find that all integrals in Eq.~\eqref{eq:first_cavity_impedance} (except the diagonal terms in Eq.~\eqref{eq:radiation_impedance_def}) evaluate $G_{0}$ or its derivatives in the far-field limit.  Thus, we approximate $G_{0}$ and its derivatives with their asymptotic forms, which replaces $\boldsymbol{K}$ with Bogomolny's transfer operator $\boldsymbol{T}$ \cite{bogolmony_defines_T}, which in the electromagnetic case is given by
\begin{equation}\label{eq:G_0_semiclassical}
    T(q,q';k)=\frac{-i}{4}\sqrt{D_{\vec{r},\vec{r}'}}e^{i S(\vec{r},\vec{r}';k)-i\pi/4}\sqrt{\frac{\cos(\theta_{f})}{\cos(\theta_{i})}},
\end{equation}
where $\theta_{i}$($\theta_{f}$) is the angle between the initial(final) wave vector and the surface at the position it leaves(hits), $S(\vec{r},\vec{r}';k)$ is the classical action along the direct trajectory from $\vec{r}$ to $\vec{r}'$, and $\sqrt{D_{\vec{r},\vec{r}'}}$ is the stability of the orbit from $\vec{r}$ to $\vec{r}'$, defined formally as
\begin{equation}\label{eq:D_def}
    D_{\vec{r},\vec{r}'}=\frac{2}{\pi k^2} \left|\frac{\partial^2 S(\vec{r},\vec{r}')}{\partial \vec{r}_{\bot}\partial \vec{r}'_{\bot}}\right|,
\end{equation}
where the derivative with respect to $\vec{r}_{\bot}$($\vec{r}_{\bot}'$) denotes the gradient with respect to $\vec{r}$($\vec{r}$) dotted into a unit vector perpendicular to the initial(final) momentum of the classical trajectory from $\vec{r}$ to $\vec{r}'$.

We note that the approximation made in Eq.~\eqref{eq:G_0_semiclassical} can be used to extend our theory beyond Eq.~\eqref{eq:Helmholtz_equation_local} by simply changing $\boldsymbol{T}$ to represent the semiclassical approximation for other physical situations.  For instance, it is possible to allow $\boldsymbol{T}$ to have a different action depending on the direction of travel, thus violating time-reversal symmetry.  It is also possible to add loss or gain to the system by adding a complex component to the action.

With these assumptions and approximations, we find
\begin{equation}\label{eq:G_u_fourier_form}
     \boldsymbol{G_{0}}u_{p}(\vec{r}')\approx  \frac{-i\pi}{2k}\sqrt{D_{\vec{r},\vec{r}'}}e^{i S(\vec{r},\vec{r}_{0,p})-i\pi/4}\tilde{u}_{p}(-\vec{k}_{i}(\vec{r},\vec{r}')),
\end{equation}
where we have expanded $S(\vec{r},\vec{r}')\approx S(\vec{r},\vec{r}_{0,p})-\vec{k}_{i}(\vec{r},\vec{r}_{0,p})\cdot\Delta \vec{r}$, the quantity $\vec{k}_{i}(\vec{r},\vec{r}')$ is the initial wave vector for the classical trajectory from $\vec{r}$ to $\vec{r}'$ , and $\tilde{u}_{p}(\vec{k})$ is the Fourier transform of $u_{p}(\vec{r})$ centered on $\vec{r}_{0,p}$:
\begin{equation}\label{eq:u_fourier_transform}
    \tilde{u}_p(\vec{k})=\int d^2\Delta\vec{r}\,e^{i \vec{k}\cdot\Delta\vec{r}}u_{p}(\vec{r}_{0,p}+\Delta \vec{r}).
\end{equation}

If we insert Eq.~\eqref{eq:fredholm_inversion} into Eq.~\eqref{eq:first_cavity_impedance}, we find
\begin{eqnarray}\label{eq:impedance_exact_sum}
    Z_{n,m}&=&Z_{R,n,m}\\&&+\frac{i k h \eta \sum_{n=0}^{\infty}\sum_{r=0}^{n} d_{r}\int  d^2\vec{r}\,u_{n}(\vec{r})\boldsymbol{V_{-}}\boldsymbol{K}^{n-r}\boldsymbol{V_{+}}u_{m}(\vec{r})}{\sum_{n=0}^{\infty}d_{n}}.\nonumber
\end{eqnarray}
Thus the impedance depends only on $d_n$, which depends on $\textrm{Tr}(\boldsymbol{K}^{n})$, and on integrals over the operators $\boldsymbol{V_{-}}\boldsymbol{K}^{n-r}\boldsymbol{V_{+}}$.  Evaluating all integrals using stationary phase, the prefactors and wave vectors are selected such that \cite{gutzwiller_book,bogolmony_defines_T}
\begin{eqnarray}\label{eq:semiclassic-sum}
    \int d^2\vec{r}\,u_{n}(\vec{r})\boldsymbol{V_{-}}\boldsymbol{K}^{l-1}\boldsymbol{V_{+}}u_{m}(\vec{r})=&\\\sum_{b(l,m,n)}\frac{\tilde{u}_{n}(\vec{k}_f) \tilde{u}_{m}^{*}(\vec{k}_i)}{4}&\sqrt{D_{b(l,m,n)}}e^{i S_{b(l,m,n)}-i\pi/4},\nonumber
\end{eqnarray}
where $b(l,m,n)$ is an index over all classical trajectories that bounce $l$ times, starting at the center of port $m$ and ending at the center of port $n$, $S_{b(l,m,n)}$ is the action for the corresponding classical trajectory and $D_{b(l,m,n)}$ is the stability coefficient defined as in Eq.~\eqref{eq:D_def} with $S(\vec{r},\vec{r}')\rightarrow S_{b(l,m,n)}$.

At this point we bring attention to the fact that the boundary of the cavity does not need to be connected.  For example, in the cavity of Fig.~\ref{fig:ensemble-picture} there is a circular perturber that is moved about, creating an ensemble of different cavities.  That circle represents a portion of the boundary that is not connected to the outer portion of the cavity boundary.  Equation~\eqref{eq:semiclassic-sum} in principle includes trajectories that pass through the perturber in going from one point on the surface to another.  (In addition Eq.~\eqref{eq:semiclassic-sum} includes trajectories that can pass through the convex upper boundary of Fig.~\ref{fig:ensemble-picture}.)  However, Bogolmony considered such trajectories \cite{bogolmony_defines_T} and found that such unphysical orbits terms come in pairs whose semiclassical contributions cancel exactly.  Thus, the sum over all semiclassical bounce terms, which is all that we will consider, will include only physical contributions.

It was found in previous work \cite{Henry_paper_one_port} that the radiation resistance for our model ports is given by
\begin{equation}\label{eq:radiation_impedance_old}
    R_{R,p}(k)=\frac{k h \eta}{4}\int\frac{d\theta}{2\pi}|\tilde{u}_{p}(k\hat{\theta})|^2,
\end{equation}
where $\hat{\theta}$ is the two-dimensional unit vector $(\cos(\theta),\sin(\theta))$,
which when inserted into Eq.~\eqref{eq:semiclassic-sum} gives
\begin{eqnarray}\label{eq:semiclassic-rad-sum}
    k h \eta\int d^2\vec{r}\,u_{n}(\vec{r})\boldsymbol{V_{-}}\boldsymbol{K}^{l-1}\boldsymbol{V_{+}}u_{m}(\vec{r})=&\\\sqrt{R_{R,n}R_{R,m}}\sum_{b(l,m,n)}C_{b(l,m,n)}&e^{i S_{b(l,m,n)}-i\pi/4},\nonumber
\end{eqnarray}
where
\begin{equation}\label{eq:C_def}
    C_{b(l,m,n)}=\frac{u_{n}(\vec{k}_{f})u_{m}^{*}(\vec{k}_{i})}{\sqrt{\langle |u_{n}|^{2}\rangle\langle|u_{m}|^{2}\rangle}}\sqrt{D_{b(l,m,n)}}.
\end{equation}
Using similar logic, we can also calculate the off-diagonal terms of Eq.~\eqref{eq:radiation_impedance_def},
\begin{equation}\label{eq:semiclassic-direct}
    Z_{R,n,m}=\sqrt{R_{R,n}R_{R,m}}C_{(0,m,n)}e^{i S_{(0,m,n)}-i\pi/4}, n\neq m,
\end{equation}
where $C_{(0,m,n)}$ and $S_{(0,m,n)}$ are the corresponding prefactor and action for a direct orbit from port $m$ to port $n$.  We view the sum over $b(l,m,n)$ as adding successively longer length orbits, and we thus refer to Eqs.~\eqref{eq:radiation_impedance_old}-\eqref{eq:semiclassic-direct} as the `short orbit formalism'.

With this result and similar semiclassical results for $\mbox{Tr}(\boldsymbol{K}^{l})$ \cite{Georgeot_prange_1995}, it is possible in principle to evaluate Eq.~\eqref{eq:first_cavity_impedance} semiclassically for any cavity.  We will not need to perform this entire calculation explicitly, however.  Instead, we will use the results from the next section to relate the sums over classical trajectories to the elements of random matrices.

We test this short-orbit formalism using the HFSS program.  In the simulator, we construct a fully three-dimensional cavity and antenna system similar to the one used in previous research; the quasi-2D nature of the cavity is enforced by choosing the excitation frequency to be below the cut-off frequency for modes that vary between the top and bottom plates.

\begin{figure}
\begin{center}
\includegraphics[width=3in]{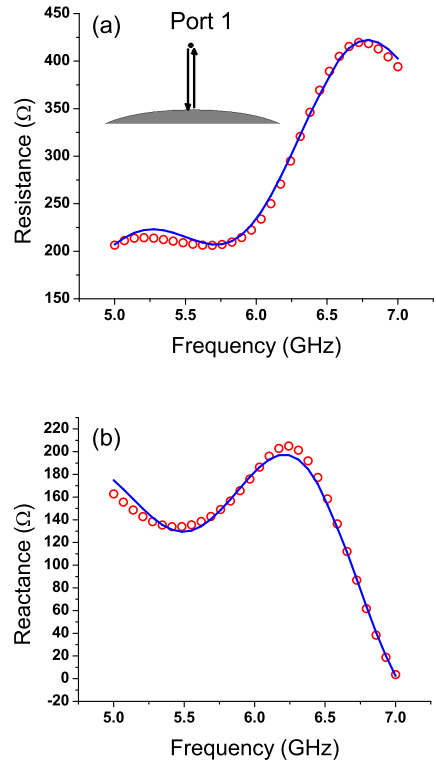}
\end{center}
\renewcommand{\baselinestretch}{1}
\begin{quote}
\caption{\label{fig:wall-A-orbit}
Comparison of the simulated impedance (circles) and theoretical impedance (solid line) of a single port with a circular perfectly conducting scatterer on one side and radiation boundary conditions on all other sides (see the inset in plot a).  The radius of curvature of the scatterer is 1.02 meters and its surface is 7.6 centimeters from the port.  The radiation impedance and phase shift for the port were extracted from independent simulation data.  Plot (a) shows the resulting resistances and plot (b) shows the reactances. (Color online)}
\end{quote}
\end{figure}

\begin{figure}
\begin{center}
\includegraphics[width=3in]{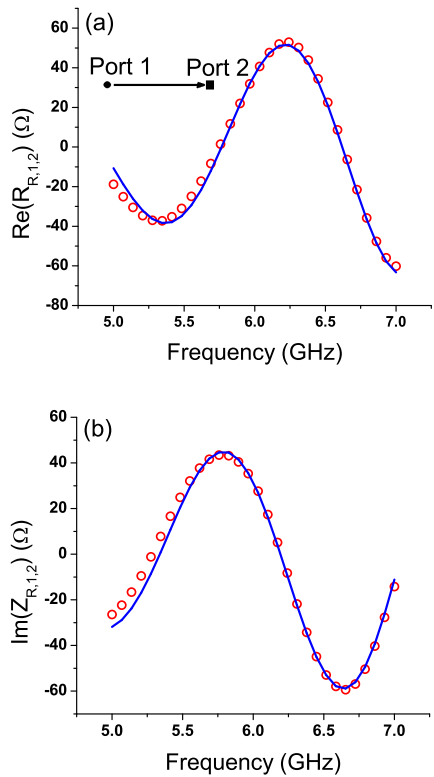}
\end{center}
\renewcommand{\baselinestretch}{1}
\begin{quote}
\caption{\label{fig:direct-orbit}
Comparison of the simulated (circles) and theoretical (solid line) $Z_{R,1,2}$ when the ports are separated by a distance of 14.4 centimeters.  Note that unlike the diagonal impedance matrix elements, the real part of the off-diagonal terms can be negative.  The radiation impedances and phase shifts were extracted from independent simulations of each port. (Color online)}
\end{quote}
\end{figure}

\begin{figure}
\begin{center}
\includegraphics[width=3in]{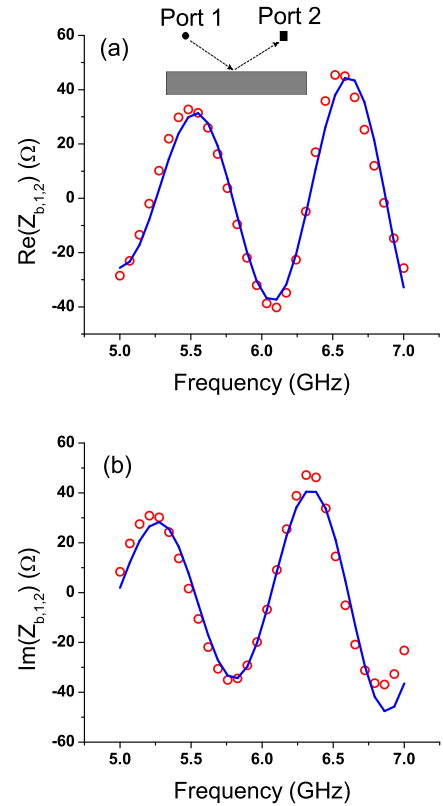}
\end{center}
\renewcommand{\baselinestretch}{1}
\begin{quote}
\caption{\label{fig:port1port2bounced}
Comparison of the simulated (circles) and theoretical (solid line) impedance due to the orbit shown in the inset in (a).  The ports were 14.4 centimeters apart and 10 centimeters from the reflecting wall.  This impedance is found by finding the total impedance of the system with the reflecting wall nearby and radiation boundary conditions everywhere else and then subtracting the radiation impedance as found in Fig.~\ref{fig:direct-orbit}. We denote it $Z_{b,1,2}$. The radiation impedances and phase shifts were extracted for each port from independent simulation data. (Color online)}
\end{quote}
\end{figure}

To test our short-orbit theory, we first simulate the radiation impedance of a single cylindrically-symmetric antenna by placing the antenna inside a circular cavity, where the outer circular wall has absorbing boundary conditions (to simulate the radiation condition of purely outgoing waves) and the port was off-center (this was to reduce coherent numerical reflections from the outer wall; the numerical absorbing boundary condition is imperfect).  As expected, we find a slowly varying function of frequency for both the radiation resistance $R_{R}$ and reactance $X_{R}$.  We then change the cavity by introducing one perfectly conducting wall into the system, effectively producing a cavity in which all waves would either radiate away or bounce once off the single wall and then radiate away, thus isolating a single term in Eq.~\eqref{eq:semiclassic-rad-sum}.  Plots of such isolated bounce terms are shown in Figs.~\ref{fig:wall-A-orbit}-\ref{fig:port1port2bounced}.

\begin{figure}
\begin{center}
\includegraphics[width=3in]{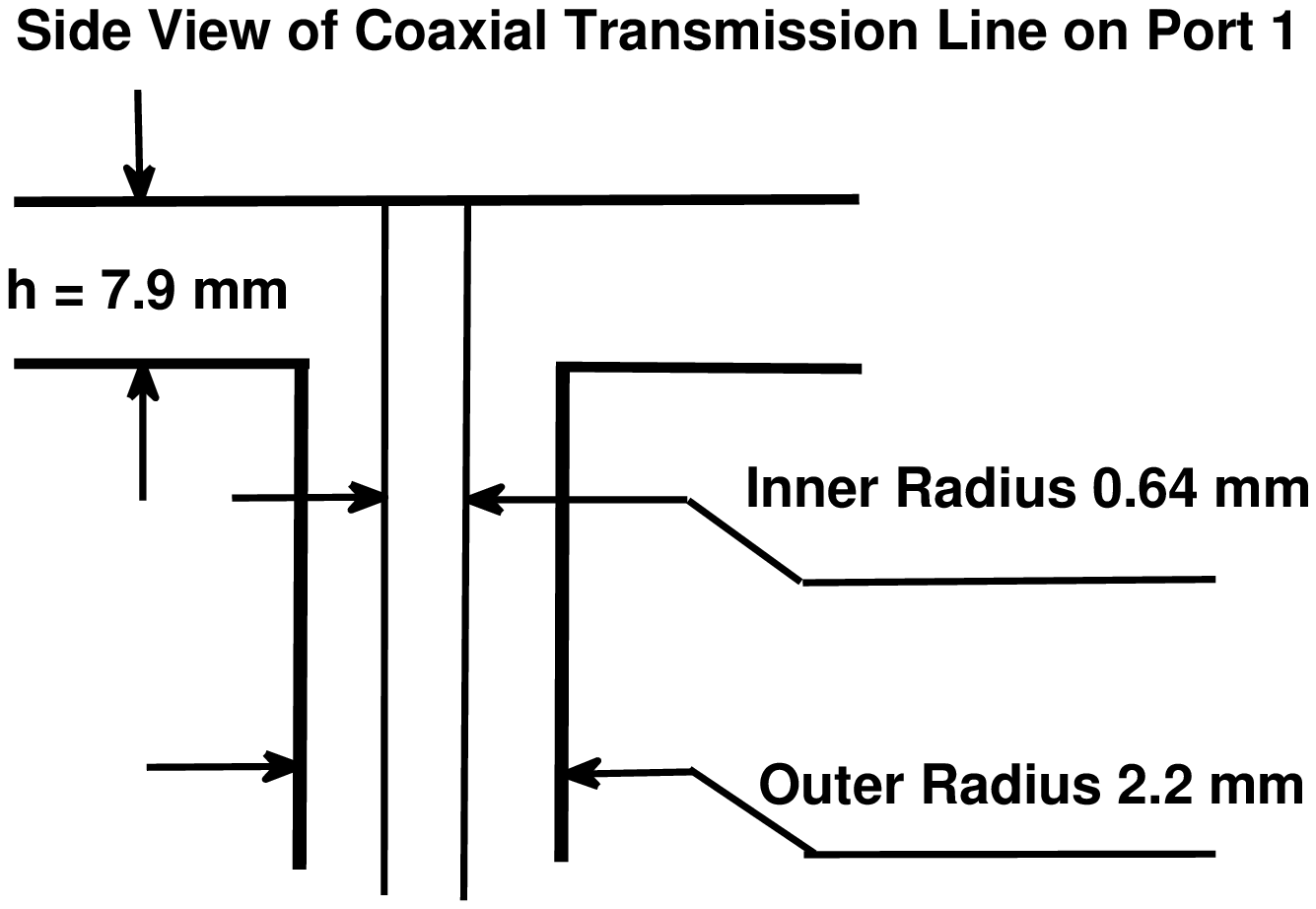}
\end{center}
\renewcommand{\baselinestretch}{1}
\begin{quote}
\caption{\label{fig:coax_cross_section}
Cross-section of the port connecting transmission line 1 and the microwave cavity.}
\end{quote}
\end{figure}

We find empirically that each short orbit term experiences a frequency-dependent phase shift $\Delta\phi_{p}$ when coupling through the port $p$, requiring the introduction of a phase factor $e^{i(\Delta\phi_n+\Delta\phi_m)}$ to Eq.~\eqref{eq:semiclassic-rad-sum}.  Although this phase shift is frequency dependent, in the cylindrically symmetric case it is orbit-independent; thus it is possible to measure the phase shift using one short orbit and then apply it to all others.  For our ports, it is most convenient to introduce this phase shift and the cylindrical symmetry by simply setting
\begin{equation}\label{eq:C_with_phase_shift}
    C_{b(l,m,n)}=e^{i(\Delta\phi_n+\Delta\phi_m)}\sqrt{D_{b(l,m,n)}}.
\end{equation}
This phase shift exists due to the fact that in the HFSS simulations, we model the ports in detail as a  circular cross-section coaxial transmission line in which the outer conductor contacts the upper plate and the inner conductor extends the short way across the cavity and contacts the lower plate.  The shape and dimensions of port 1 are shown in Fig.~\ref{fig:coax_cross_section}.  Port 2, when it is present, has the same geometry as port 1, but with an outer radius of 3.0 mm.  This more detailed port model results in an additional phase shift that is not treated in our simple model of Eq.~\eqref{eq:Helmholtz_equation_local} where we add a fixed current source to the wave equation.  With this phase correction in $C_{b(l,m,n)}$, however, we find that we can model the impedance very well by using Eq.~\eqref{eq:semiclassic-rad-sum}.

To show the agreement between theory and simulation, we create three different cavity configurations.  In the first configuration, we have a single port with a single conducting, curved wall near it, with absorbing boundary conditions on the remaining surfaces.  After simulating the isolated ports radiation impedance and finding the phase shift, we both predict and simulate the resulting impedance in the presence of the curved wall.  Fig.~\ref{fig:wall-A-orbit} shows the measured and the predicted impedance from 5 to 7 GHz, both resistance and reactance, where it can be seen that the theoretical and simulated results agree well.  The deviations between the two are expected; the semiclassical approximation is not perfect at frequencies this low, and we also have diffraction from the necessary truncation of the wall.

In the second configuration, we tested Eq.~\eqref{eq:semiclassic-sum} by introducing a second antenna into the system and imposing radiation boundary conditions on all the outer walls.   We found the radiation impedance and phase shift of the second antenna using exactly the same methods as for the first.  We then simulated the mutual impedance between the two ports and compared it to Eq.~\eqref{eq:semiclassic-direct} using the simulated port parameters, and found that the agreement was again excellent, as shown in Fig~\ref{fig:direct-orbit}.

The third configuration is the same as the second, except we add a conducting wall next to the two ports, creating an orbit which leaves the first port, bounces off the wall once, and goes to the second port.  To isolate the impedance due to this new orbit, we compare the changes in impedance (which we denote $Z_{b,1,2}$) between configurations 2 and 3 rather than the raw impedances.  Comparing this difference with the semiclassical prediction from Eq.~\eqref{eq:semiclassic-rad-sum}, we again see excellent agreement, as shown in Fig.~\ref{fig:port1port2bounced}.

Thus we believe that our short-orbit formulation is effective at predicting the impedance of cavities with a few, short orbits.  In the next section, we discuss an equivalent formulation which expresses the impedance as elements of a finite-dimensional matrix.  This equivalence between the matrix and semiclassical formulations will allow us to create ensembles of cavities which account for short orbits within the cavity.

\subsection{The finite matrix formulation}
\label{subsec:finite_matrix}

The results of Sec.~\ref{subsec:short-orbit-approx} are based on an evaluation of the continuous, integral operator $\boldsymbol{K}$ defined in Eq.~\eqref{eq:operator_defs}.  To make connection with random matrix theory, we wish to recast the equations in matrix form.  The authors of Ref.~\cite{fredholm_method_for_scars} have shown how to do this.  Further, in the semiclassical limit, the resulting matrix is effectively finite.

To derive the matrix formulation, we first replace the continuous operator $\boldsymbol{K}$ with the semiclassical operator $\boldsymbol{T}$.  Then we use the result that semiclassically \cite{fredholm_method_for_scars},
\begin{equation}\label{eq:v_-_v_+_relationship}
    i\boldsymbol{V_{-}}\boldsymbol{T}^{\dag}=\boldsymbol{V_{+}}^{\dag}.
\end{equation}
The operator $\boldsymbol{V_{+}}^{\dag}\boldsymbol{V_{+}}$  was also found in the semiclassical approximation to be \cite{fredholm_method_for_scars}
\begin{equation}\label{eq:bigV_+_mag}
    \boldsymbol{V_{+}}^{\dag}\boldsymbol{V_{+}} = i \left(\boldsymbol{G_{0}}-\boldsymbol{G_{0}}^{\dag}\right).
\end{equation}
By using Eq.~\eqref{eq:v_-_v_+_relationship} to eliminate $\boldsymbol{V_{-}}$ and by adding $\left[\boldsymbol{V_{+}}^{\dag}\boldsymbol{V_{+}} - i \left(\boldsymbol{G_{0}}-\boldsymbol{G_{0}}^{\dag}\right)\right]/2$, which is zero, to the operator inside Eq.~\eqref{eq:first_cavity_impedance}, we can rewrite the impedance as
\begin{eqnarray}
    Z_{n,m}=&\frac{1}{2}k h \eta\int d^2\vec{r}\,u_{n}(\vec{r})&\left[\boldsymbol{V_{+}}^{\dag}\frac{\boldsymbol{1}+\boldsymbol{T}}{ \boldsymbol{1}-\boldsymbol{T}}\boldsymbol{V_{+}}\right.
    \\&&\left.+i\left(\boldsymbol{G_{0}}+\boldsymbol{G_{0}}^{\dag}\right)\vphantom{\frac{\boldsymbol{1}}{\boldsymbol{1}}}\right]u_{m}(\vec{r}),\nonumber\\
    =&iX_{R,n,m}+\frac{1}{2}k h \eta & \int d^2\vec{r}\,u_{n}(\vec{r})\boldsymbol{V_{+}}^{\dag}\frac{\boldsymbol{1}+\boldsymbol{T}}{ \boldsymbol{1}-\boldsymbol{T}}\boldsymbol{V_{+}}u_{m}(\vec{r}).\nonumber
\end{eqnarray}

Fishman, Prange and Georgeot \cite{fredholm_method_for_scars} demonstrated that in the semiclassical limit, the operator $\boldsymbol{T}$  can be represented as an infinite-dimensional matrix whose components are zero except on a finite subspace of dimension $N=2L/\lambda$, where in our formulation $L$ is the circumference of the cavity.  They demonstrated this by expanding all functions on their surface of section (in our formulation, the cavity boundary) in a Fourier series. In this basis, an arbitrary function $\nu(q)$ is expanded as
\begin{equation}\label{eq:surface_expansion}
    \nu(q)=\sum_{n=-\infty}^{\infty}a_{n} e^{2\pi n i q/L}.
\end{equation}
They showed that the operator $\boldsymbol{T}$, evaluated using stationary phase in this basis, is insensitive to Fourier components smaller than a wavelength, resulting in the truncated subspace.  By identical logic, the operator $\boldsymbol{V_{+}}$ only projects onto this semiclassical subspace.  Thus, semiclassically, the function $\boldsymbol{V_{+}}u_{p}(\vec{r})$ is non-zero only on this subspace, where it has $N$ discrete components corresponding to the Fourier components of the expansion in Eq.~\eqref{eq:surface_expansion}.

Using Eq.~\eqref{eq:bigV_+_mag}, we get the dot product between two of these vectors,
\begin{eqnarray}
    \int d^2\vec{r}\,u_{n}(\vec{r})\boldsymbol{V_{+}}^{\dag}\boldsymbol{V_{+}}u_{m}(\vec{r})&=&i \int d^2\vec{r}\,u_{n}(\vec{r})\left(\boldsymbol{G_{0}}-\boldsymbol{G_{0}}^{\dag}\right)u_{m}(\vec{r})\nonumber\\
    &=&\frac{2\tilde{R}_{R,n,m}}{k h \eta}.\label{eq:little_v_mag_def}
\end{eqnarray}
Thus we can rewrite Eq.~\eqref{eq:second_cavity_impedance} as
\begin{equation}\label{eq:matrix_cavity_impedance}
     \boldsymbol{Z}=i\boldsymbol{\tilde{X}}_{R}+\boldsymbol{v}^{\dag}\cdot\frac{\boldsymbol{1}+\boldsymbol{T}}{ \boldsymbol{1}-\boldsymbol{T}}\cdot\boldsymbol{v}
\end{equation}
where we now treat $\boldsymbol{T}$ as an $N\times N$ matrix and where $\boldsymbol{v}$ is an $N\times M$ matrix whose columns $\vec{v}_p$ are the $N$-dimensional vectors proportional to the semiclassical $\boldsymbol{V_{+}}u_{p}(\vec{r})$ and which are normalized such that
\begin{equation}\label{eq:littleV_mag}
    \boldsymbol{v}^{\dag}\boldsymbol{v}=\boldsymbol{\tilde{R}}_{R}.
\end{equation}

Because Eqs.~\eqref{eq:semiclassic-rad-sum}, \eqref{eq:semiclassic-direct} and \eqref{eq:littleV_mag} are all evaluated in the stationary phase approximation, we can equate the matrix elements of any power of $\boldsymbol{T}$ with the semiclassical orbit terms
\begin{equation}\label{eq:orbit-bounced}
    2\vec{v}_{n}^{\dag}\cdot \boldsymbol{T}^{l-1}\cdot\vec{v}_{m}=\sqrt{R_{R,n}R_{R,m}}\sum_{b(l,m,n)}C_{b(l,m,n)}e^{i S_{b(l,m,n)}-i\pi/4}.
\end{equation}

Equation~\eqref{eq:orbit-bounced} is one of the most important results of this paper.  By explicitly connecting the semiclassical sums to a semiclassical matrix formulation of impedance, we can relate the classical trajectories to the more abstract operator formalism.  Thus when we create ensembles of $\boldsymbol{v}$ and $\boldsymbol{T}$, we can relate the ensemble averages of $\boldsymbol{v}$ and $\boldsymbol{T}$ to the corresponding ensemble averages of the classical trajectories within the cavity, which gives us a natural method of creating ensembles of $\boldsymbol{T}$ which are constrained by short orbits within the system.

\section{Impedance Statistics}\label{sec:statistics}

As noted in the introduction, it is often difficult to solve the wave equation exactly.  Even in the semiclassical regime, where the problem is in principle tractable using classical trajectories, there are difficulties.  If the classical dynamics is chaotic, the number of classical trajectories grows exponentially as does their sensitivity to numerical errors. Small mistakes in modeling or small changes between similar systems will result in large changes in the observed behavior.  Thus we follow the long-standing tradition of replacing our deterministic expressions with statistical models which reproduce the generic behavior of the systems being considered.

Our model for the cavity impedance is given in Eq.~\eqref{eq:matrix_cavity_impedance}. The matrices $\boldsymbol{\tilde{X}}_{R}$ and $\boldsymbol{v}$ (up to an unmeasurable and thus arbitrary basis, which can thus be absorbed into $\boldsymbol{T}$) are determined by Eqs.~\eqref{eq:radiation_impedance_def} and \eqref{eq:radiation_reactance_resistance}, which depend only on the radiation fields from the ports, and are thus amenable to direct measurement or non-chaotic semiclassical theory. Thus to find our statistical properties, we simply seek an appropriate distribution for $\boldsymbol{T}$.

The matrix $\boldsymbol{T}$ may be viewed as representing an internal scattering matrix.  In the case of chaotic dynamics and in the context of random matrix theory, it is most natural to model $\boldsymbol{T}$ as an element of Dyson's circular ensemble \cite{Dyson_original} (with the time-reversal symmetry determined by the symmetry of the underlying system).  If we make this substitution, it can be shown that the resulting statistical properties of $\boldsymbol{Z}$ are completely equivalent to our previously published random coupling model.  In this model, the system specific properties of the ports (specifically the radiation impedance) are all that is used to ``normalize'' the statistically fluctuating impedance.  Effectively, the previous model assumes that once wave energy enters the cavity it is randomized by the chaotic ray trajectories such that no details of the interior of the cavity modify the statistics of the impedance.  However, we have seen in Fig.~\ref{fig:median-deviation} that there is likely some influence of specific ray trajectories within the cavity on the statistical properties (in the case of Fig.~\ref{fig:median-deviation} the median) of the impedance.  We now assume that the effect of these trajectories can described by the Poisson kernel \cite{Poisson_Kernel_Original}. That is, we assume that the distribution of $\boldsymbol{T}$ is given by the Poisson kernel \cite{Poisson_including_internal}
\begin{equation}\label{eq:Poisson_kernel}
    P(\boldsymbol{T})=\frac{1}{2^{N(\beta N+2-\beta)/2} V}\frac{\det\left(\boldsymbol{1}-\boldsymbol{\bar{T}}^{\dag}\boldsymbol{\bar{T}}\right)^{(\beta N+2-\beta)/2}}{\det\left(\boldsymbol{1}-\boldsymbol{\bar{T}}^{\dag}\boldsymbol{T}\right)^{\beta N+2-\beta}}
\end{equation}
where $V$ is a normalization constant and $\boldsymbol{\bar{T}}$ is the average value of $\boldsymbol{T}$ over the ensemble, and $\boldsymbol{\bar{T}}$ is, in principle, determined explicitly by the boundaries of the cavities in the ensemble.

The Poisson kernel does not just specify the average value of $\boldsymbol{T}$; it has the general property that \cite{Poisson_Kernel_Original}
\begin{equation}\label{eq:T_l_avg}
    \langle \boldsymbol{T}^{l}\rangle = \boldsymbol{\bar{T}}^{l}.
\end{equation}
If we knew $\boldsymbol{\bar{T}}$, then we could find the distribution of $\boldsymbol{Z}$ directly.  Unfortunately, finding $\boldsymbol{\bar{T}}$ for a specific ensemble such as that shown in Fig.~\ref{fig:ensemble-picture} is almost as complex as finding $\boldsymbol{T}$ and thus has no advantage over numerically solving the Helmholtz equation.  By averaging both sides of Eq.~\eqref{eq:orbit-bounced}, however, we can find the components of $\boldsymbol{\bar{T}}^{l}$ spanned by the column vectors of $\boldsymbol{v}$.  We find that knowing these average short orbit terms for all $l$ is sufficient to get the statistics of $\boldsymbol{Z}$; because the sum over average short orbits is expected to converge, the problem becomes tractable for a wide range of ensembles.

With this assumption for the distribution of $\boldsymbol{T}$, we can find the statistical properties of $\boldsymbol{Z}$.  From a result due to Brouwer for matrices distributed according to the Poisson kernel \cite{Brouwer_Lorentzian}, we find that we can parameterize $\boldsymbol{T}$ as
\begin{equation}\label{eq:Lorentzian_scattering}
    \boldsymbol{T}=\frac{i\boldsymbol{W}^{\dag}\left(\lambda \boldsymbol{\tilde{H}_0}+\epsilon\boldsymbol{1}\right)\boldsymbol{W}-\boldsymbol{1}}{i\boldsymbol{W}^{\dag}\left(\lambda \boldsymbol{\tilde{H}_0}+\epsilon\boldsymbol{1}\right)\boldsymbol{W}+\boldsymbol{1}},
\end{equation}
where the scalars $\lambda$, $\epsilon$ and the $N\times N$ matrix $\boldsymbol{W}$ are ensemble-specific constants which depend only on $\boldsymbol{\bar{T}}$, and $\boldsymbol{\tilde{H}_0}$ is an $N\times N$ random matrix distributed according to the pdf of the Lorentzian ensemble with median 0 and width 1,
\begin{equation}\label{eq:Lorentzian_pdf}
    P(\boldsymbol{\tilde{H}_{0}})=\frac{1}{V}\frac{\lambda^{N(\beta N+2-\beta)/2}}{\det(\boldsymbol{1}+\boldsymbol{\tilde{H}}^2)^{(\beta N+2-\beta)/2}},
\end{equation}
where $\beta=1(2,4)$ for the Orthogonal(Unitary,Symplectic) choice of time-reversal behavior.
Inserting Eq.~\eqref{eq:Lorentzian_scattering} into Eq.~\eqref{eq:matrix_cavity_impedance}, we find
\begin{equation}\label{eq:impedance_function_of_H_0}
    \boldsymbol{Z}=i\boldsymbol{\tilde{X}}_{R}+i\lambda(\boldsymbol{W}\boldsymbol{v})^{\dag}\boldsymbol{\tilde{H}}_{0}\boldsymbol{W}\boldsymbol{v}+i\epsilon(\boldsymbol{W}\boldsymbol{v})^{\dag}\boldsymbol{W}\boldsymbol{v}.
\end{equation}

We now wish to eliminate $\lambda$, $\epsilon$ and $\boldsymbol{W}\boldsymbol{v}$ from Eq.~\eqref{eq:impedance_function_of_H_0}.  We do this by noting that given the parametrization in Eq.~\eqref{eq:Lorentzian_scattering}, Brouwer found the value of $\boldsymbol{\bar{T}}$ to be \cite{Brouwer_Lorentzian}
\begin{equation}\label{eq:Lorentzian_poisson_average}
    \boldsymbol{\bar{T}}=\frac{(\lambda+i\epsilon)\boldsymbol{W}^{\dag}\boldsymbol{W}-\boldsymbol{1}}{(\lambda+i\epsilon)\boldsymbol{W}^{\dag}\boldsymbol{W}+\boldsymbol{1}}.
\end{equation}
Solving Eq.~\eqref{eq:Lorentzian_poisson_average} for $\boldsymbol{W}^{\dag}\boldsymbol{W}$ and projecting both sides onto the subspace spanned by $\boldsymbol{v}$, we get
\begin{equation}\label{eq:lambda_epsilon_Wv_solve}
    (\lambda+i\epsilon)\left(\boldsymbol{W}\boldsymbol{v}\right)^{\dag}\cdot\left(\boldsymbol{W}\boldsymbol{v}\right)=\boldsymbol{Z}_{avg}-i\boldsymbol{\tilde{X}}_{R},
\end{equation}
where we formally define $\boldsymbol{Z}_{avg}$ as
\begin{equation}\label{eq:Z_avg}
    \boldsymbol{Z}_{avg}=i\boldsymbol{\tilde{X}}_{R}+\boldsymbol{v}^{\dag}\cdot\frac{\boldsymbol{1}+\boldsymbol{\bar{T}}}{\boldsymbol{1}-\boldsymbol{\bar{T}}}\cdot\boldsymbol{v}.
\end{equation}
We denote the Hermitian and anti-Hermitian components of $\boldsymbol{Z}_{avg}$ as $\boldsymbol{R}_{avg}$ and $\boldsymbol{X}_{avg}$, respectively.  Matching the Hermitian and anti-Hermitian components across the equality in Eq.~\eqref{eq:lambda_epsilon_Wv_solve} and noting that $\boldsymbol{R}_{avg}$ must be a non-negative matrix because $\boldsymbol{\bar{T}}$ is subunitary and normal, we find that
\begin{eqnarray}\label{eq:epsilon_solve}
  \lambda \left(\boldsymbol{W}\boldsymbol{v}\right)^{\dag}\boldsymbol{W}\boldsymbol{v}& = &\boldsymbol{R}_{avg}\\\label{eq:lambda_solve}
  \epsilon \left(\boldsymbol{W}\boldsymbol{v}\right)^{\dag}\boldsymbol{W}\boldsymbol{v}& = & \boldsymbol{X}_{avg} - \boldsymbol{\tilde{X}}_{R}.
\end{eqnarray}
Inserting these results into Eq.~\eqref{eq:impedance_function_of_H_0} gives us
\begin{equation}\label{eq:impedance_final}
    \boldsymbol{Z}=i\boldsymbol{X}_{avg}+i\sqrt{\boldsymbol{R}_{avg}}\boldsymbol{\xi}\sqrt{\boldsymbol{R}_{avg}}
\end{equation}
where $\boldsymbol{\xi}$ is $\boldsymbol{\tilde{H}}_0$ projected onto the subspace spanned by $\boldsymbol{W}\boldsymbol{v}$.  Brouwer proved that any diagonal submatrix of a Lorentzian distributed matrix is also a Lorentzian distributed matrix with the same median and width.  This result combined with the basis invariance of Eq.~\eqref{eq:Lorentzian_pdf} leads to the conclusion that $\boldsymbol{\xi}$ is a Lorentzian random matrix with width 1 and median 0, as predicted.  The basis-invariance of the distribution of $\boldsymbol{\tilde{H}}_{0}$ also means that although Eq.~\eqref{eq:epsilon_solve} has a family of related solutions for $\boldsymbol{W}\boldsymbol{v}$, all members of this family are related via a change of basis and therefore result in identical statistics for $\boldsymbol{Z}$.

Although knowing the form of the distribution for $\boldsymbol{Z}$ is useful and can be used fruitfully to fit experimental or simulation data, at a single frequency it is only a minor improvement over the Poisson kernel in which one can also extract $\boldsymbol{\bar{S}}$ from numerical data \cite{Poisson_including_internal,Poisson_including_internal_2}. Our last step is therefore to predict the value of $\boldsymbol{Z}_{avg}$ using the semiclassical approximations developed in Sec.~\ref{subsec:short-orbit-approx}.  We do this by noting that because $\boldsymbol{\bar{T}}$ is subunitary, the magnitude of all its eigenvalues are less than or equal to one.  The set of $\boldsymbol{\bar{T}}$ which have any eigenvalues on the unit circle has measure zero.  Therefore we can expand Eq.~\eqref{eq:Z_avg} in a convergent series as
\begin{equation}\label{eq:Z_avg_expansion}
    \boldsymbol{Z}_{avg}=\boldsymbol{\tilde{Z}}_{R}+2\sum_{l=1}^{\infty}\boldsymbol{v}^{\dag}\cdot \boldsymbol{\bar{T}}^{l}\cdot\boldsymbol{v}.
\end{equation}
Substituting Eqs.~\eqref{eq:orbit-bounced} and \eqref{eq:semiclassic-direct} into Eq.~\eqref{eq:Z_avg_expansion}, and remembering Eq.~\eqref{eq:T_l_avg}, we see that semiclassically $\boldsymbol{Z}_{avg}$ is the port impedance plus the sum of the average contributions the short orbits make to the impedance.  In our case, where we have a perturber which moves much more than a wavelength between realizations through the entire cavity,  the contributions of orbits reflected off the perturber will have an essentially random phase, and we approximate their contributions as zero.  Because the outer walls are fixed, short orbits from the ports to the walls will systematically appear in the sum in Eq.~\eqref{eq:Z_avg_expansion}, but must be weighted by $p_{b}$, the fraction of realizations in which they do not pass through the perturber.  With these results, we find the elements of $\boldsymbol{\zeta}$ from Eq.~\eqref{eq:Z_avg_introduction} to be
\begin{equation}\label{eq:zeta_def}
    \zeta_{n,m}=\sum_{b(n,m)}p_{b(n,m)}C_{b(n,m)}e^{i S_{b(n,m)}(k)-i\pi/4}
\end{equation}
where the index $b(n,m)$ is over all short orbits which go from port $m$ to port $n$, including direct orbits between different ports.  Note that when we test this theory, we use the empirically discovered form of $C_{b(n,m)}$ from Eq.~\eqref{eq:C_with_phase_shift}.

We note that $\boldsymbol{Z}_{avg}$ is the impedance the baseline system would have if some fraction of energy were lost every time a wave passed through a perturber.  Even for very large numbers of bounces, this seems to be a general result:  the impedance needed to normalize the statistics of any sufficiently random ensemble will correspond to the impedance of a single lossy cavity where loss occurs in those features which change between realizations, with the degree of loss determined by the degree of change in those elements.  Thus even with very small perturbations in which the semiclassical approach is unfeasible, the form of $\boldsymbol{Z}_{avg}$ is known, and in analogy to the Poisson kernel, we can fit to find the effective radiation impedance.  More importantly this implies that the frequency dependence of $\boldsymbol{Z}_{avg}$ matches that of an appropriate lossy cavity.  If the perturbations are sufficiently uniform within the cavity, the statistics of $\boldsymbol{Z}_{avg}$ evaluated over a range of sufficiently separate frequencies would exhibit the statistics found in our previous work for lossy cavities.

In microwave billiards exhibiting hard chaos with a uniform distribution of perturber locations within the volume, it is possible to estimate the expected loss parameter for $\boldsymbol{Z}_{avg}$.  Because typical ray trajectories within the cavity explore the phase space ergodically and because the perturber locations are distributed uniformly, we expect that, $p_{b(n,m)}\sim \exp(-\tilde{\alpha} L_{b(n,m)})$, where $\tilde{\alpha}$ is determined by the perturber size and shape and the perturber locations, and $L_{b(n,m)}$ is the length of the $b(n,m)$th orbit.  Neglecting the phase shifts from the traversals of the ports (which are expected to be small due to the small size of the ports), we find that with this expression for $p_{b(n,m)}$,
\begin{equation}
\zeta_{n,m}\sim \sum_{b(n,m)}C_{b(n,m)}\exp(i (k+i\tilde{\alpha}) L_{b(n,m)}-i\pi/4),
\end{equation}
where we have used that the classical action in microwave billiards is given by $S_{b(n,m)}=k L_{b(n,m)}$.  We found in previous work \cite{Henry_paper_one_port} that the transformation $k\rightarrow k+i\tilde{\alpha}$ corresponds to adding loss, with the loss parameter given by $Q=k/(2\tilde{\alpha})$ and the line-width to level spacing ratio $\alpha=k A \tilde{\alpha}/\pi$, where $A$ is the area of the microwave cavity.  Thus, by analogy, introducing $p_{b}$ into Eq.~\eqref{eq:zeta_def} is roughly equivalent to $\boldsymbol{Z}_{avg}$ being the impedance of a lossy cavity with uniform loss. Using a Monte Carlo simulation of long ray trajectories in our bowtie billiard, we find that for the perturber positions in Fig.~\ref{fig:ensemble-picture}, $\tilde{\alpha}\sim .25 m^{-1}$.

This result for the average impedance is related to work done by Brouwer and Beenakker \cite{Brouwer_and_Beenakker} following B\"{u}ttiker \cite{Buttiker_probe}.  In their work, they found that a lossy quantum dot could be modeled as a lossless quantum dot coupled both to the physical scattering channels and to a large number of weakly coupled parasitic channels.  Because $\boldsymbol{Z}_{avg}$ represents the impedance of a system where a small amount of energy is lost when it passes through a perturber position, the perturber positions function as effective parasitic channels, and we expect the resulting average impedance and average scattering matrix to be well-fit by their theory.

\section{Numerical Tests of the Theoretical Predictions}
\label{sec:numerical-testing}

In this section, we show the results of several tests of Eqs.~\eqref{eq:Z_avg_introduction}~and~\eqref{eq:zeta_def} for both one and two-port configurations.  We obtained the data for these tests from simulations using HFSS.  The cavity configuration we used is shown in Fig.~\ref{fig:ensemble-picture}.  Port 1 is centered at location ($x=18.03$cm, $y=15.48$cm) and port 2 (when it is present) is at ($x=36.7$cm,$y=15.48$cm).  The lower-left corner of the cavity is at ($x=0.0$cm, $y=0.0$cm).  Both ports have essentially the geometry as shown in Fig.~1 of Reference~\cite{Henry_paper_one_port}, but with different dimensions.  Both ports have an inner radius of .635mm, but port 1 has an outer radius of 2.29mm while port 2 has an outer radius of 3.05mm.  The lower and left straight sides of the cavity have lengths $L_{1}=43.18$cm and $L_{2}=21.59$cm respectively and the upper and right sides have radius of curvature $R_{1}=103$cm and $R_{2}=63.9$cm respectively.  By moving a perfectly conducting circular perturber with a diameter of 2.54cm to 95 different locations (shown as circles in Fig.~\ref{fig:ensemble-picture}) within the cavity, we construct our ensemble.  The resulting impedances were simulated at 201 uniformly spaced frequencies from 5 to 7 GHz, inclusive.  For this frequency range, we get that the effective loss parameter $Q$ due to the ensemble averaging changes linearly from about $210$ at 5 GHz to $300$ at 7 GHz.  The effective line-width to level spacing ratio $\alpha$ also increases linearly from about $0.95$ to $1.35$.

\subsection{Single-port Tests}
\label{subsec:one_port_stats}

To test our predictions in the single-port case, we first confirm that at each frequency the impedances have a Lorentzian distribution, fitting to find the median and width.  From the fit, we see that Eq.~\eqref{eq:zeta_def} does not converge quickly enough to be practical at a single frequency.  However, we find that if we use frequency averaging and short orbits together, we regain universal statistics over a much narrower frequency band than was required with out previous theory \cite{Henry_paper_one_port} (i.e.\ if $\boldsymbol{Z}_{R}$ is used as in Eq.~\eqref{eq:chaotic_impedance_statistics}).  In addition, we confirm that as a function of frequency, the fitted $\boldsymbol{Z}_{avg}$ has the characteristic behavior of the impedance of a lossy cavity.

To test that the impedance is Lorentzian distributed at each frequency, we numerically find the three quartiles $Q_{(1,2,3)}(f)$ of the 95 sample impedances (denoted $Z_i(f)$) and thus find the sample median $Z_{med}(f)=Q_2(f)$ and the sample width $Z_{wid}(f)=Q_3(f)-Q_1(f)$.  Assuming that $Z_{med}(f)$ and $Z_{wid}(f)$ are approximately the correct median and width, we can then find the phase of the normalized scattering matrix coefficient
\begin{equation}\label{eq:normed_sample_dist}
    \phi_{i}(f)=2\tan^{-1}\left(\frac{Z(f)-Z_{med}(f)}{Z_{wid}(f)}\right).
\end{equation}
For large $N$ and with a Lorenzian distribution for $Z(f)$, $\phi_{i}(f)$ should be uniformly distributed between $-\pi$ and $\pi$.  We bin the numerical $\phi_{i}$ into 10 equal-sized bins and find the resulting $\chi^2$ deviation from a uniform distribution.  We then find the anticipated distribution of $\chi^2$ by performing a Monte-Carlo statistical analysis on 402000 realizations of 95 Lorentzian random variables each, transforming them precisely as done in Eq.~\eqref{eq:normed_sample_dist} and finding the resulting $\chi^2$.  Comparing the values of $\chi^2$ for the simulation data and the Monte-Carlo distribution, we find that at the 95\% confidence level, we can accept the Lorentzian hypothesis for 93\% of our frequency impedance samples, while at the 99\% confidence level, we can accept all of our impedance samples.  These results are consistent with the data being distributed with a Lorentzian distribution at each frequency.

To test the semiclassical theory in the single-port case, rather than using the sample median and width we attempt to predict the median and width using Eqs.~\eqref{eq:Z_avg_introduction} and \eqref{eq:zeta_def}.  In practice, we must eventually truncate the sum over semiclassical trajectories.  We therefore define the truncated average impedance
\begin{equation}\label{eq:Z_avg_truncated}
    \boldsymbol{Z}_{t,N_{b}}=\boldsymbol{\tilde{Z}}_{R}+2\sum_{l=1}^{N_{b}}\vec{v}_{1}^{\dag}\cdot\boldsymbol{\bar{T}}^{l}\cdot\vec{v}_{1},
\end{equation}
which is equivalent to the sum over all classical trajectories which bounce up to $N_{b}$ times.  Thus the normalized scattering phase we use to test our semiclassical theory is
\begin{equation}\label{eq:normed_theory_dist}
    \phi_{i,N_{b}}(f)=2\tan^{-1}\left(-i\frac{Z_{i}(f)-iX_{t,N_{b}}(f)}{R_{t,N_{b}}(f)}\right),
\end{equation}
where $R_{t,N_{b}}$ and $X_{t,N_{b}}$ are the real and imaginary parts of $Z_{t,N_{b}}$.

\begin{figure}
\begin{center}
\includegraphics[width=3in]{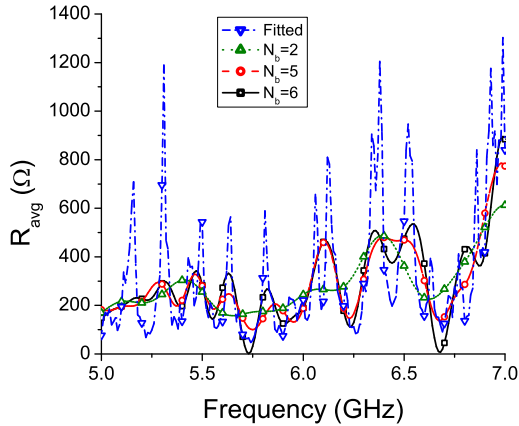}
\end{center}
\renewcommand{\baselinestretch}{1}
\small\normalsize
\begin{quote}
\caption{\label{fig:fit_vs_bounces}A comparison between the fitted width of the impedance distribution $Z_{wid}$ at each measurement frequency and the semiclassical predictions for the widths due to short orbits which bounce up to 2, 5 or 6 times.  By 6 bounces, the semiclassical prediction has begun to fit the gross features of the fitted widths but is far short of the number of orbits needed to fit the sharp spikes.  In addition, Gibbs phenomenon has become a problem in the sixth bounce around 5.75 and 6.7 GHz with the semiclassical prediction dipping too close to zero. (Color online)}
\end{quote}
\end{figure}

In Fig.~\ref{fig:fit_vs_bounces}, we compare $Z_{t,N_{b}}$ to $Z_{wid}$ for $N_{b}=2,5,6$.  We see that as $N_{b}$ increases, the two terms becomes increasingly similar.  $Z_{t,2}$ follows the frequency average of $Z_{wid}$.  Also, despite the fact that $Z_{wid}$ changes rapidly in frequency, consistent with the assumption that it is the real part of a lossy impedance, $Z_{t,5}$ and $Z_{t,6}$ have begun to fit even the large spikes in $Z_{wid}$.  This strongly supports the validity of Eq.~\eqref{eq:zeta_def} and implies that with a sufficiently large number of bounces or a more random ensemble (such as including more perturbers or having a larger perturber), it may be possible to predict $\boldsymbol{Z}_{avg}$ at a single frequency.  We have not yet confirmed this possibility.

In previous work, we found that normalizing the impedance with zero bounces (i.e.\ the radiation impedance $\boldsymbol{Z}_{R}$) was sufficient to get universal statistics if we sampled the impedances over a sufficiently wide frequency range.  If the frequency window was too narrow, however, we found systematic deviations from universal statistics. Even though we do not have enough terms in $\boldsymbol{Z}_{t,6}$ to experimentally find $\boldsymbol{Z}_{avg}$ semiclassically at a single frequency, we find that by combining short orbits and frequency averaging, we can get universal statistics over much narrower frequency ranges than previously.

To measure the deviation of the distribution of the measured $\phi_{i,N_{n}}$ from uniform, we introduce the $\chi^2$ statistic.  For a frequency window of width $\delta f$ and centered at $f_0$, the $\chi^2$ statistics is calculated by binning the $\phi_{i,N_{b}}$ from every realization and from every other frequency in the frequency window into ten equally sized bins from $-\pi$ to $\pi$.  (We take every other frequency window because the impedance values of adjacent frequencies are found to be strongly correlated.)  The $\chi^2$ statistic for this window is then given by
\begin{equation}\label{eq:chi_2_def}
    \chi^2=\sum_{r=1}^{10}\frac{(N_{r}-\langle N_{r}\rangle)^2}{\langle N_{r}\rangle}
\end{equation}
where $N_{r}$ is the number of $\phi_{i,N_b}$ in the $r$th bin, and $\langle N_{r}\rangle$ is the expected value of $N_{r}$ given a uniform distribution.  The $\chi^2$ statistic is chosen because it has approximately the same distribution independent of $\Delta f$.

\begin{figure}
\begin{center}
\includegraphics[width=3in]{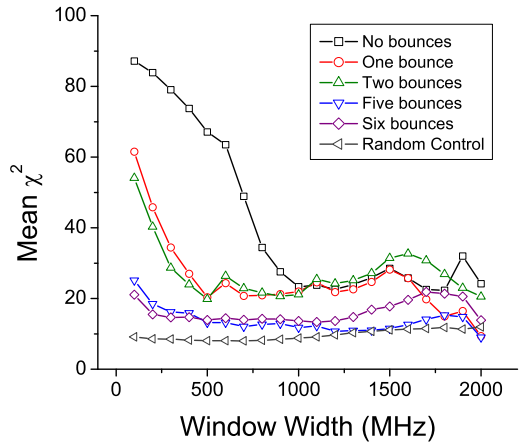}
\end{center}
\renewcommand{\baselinestretch}{1}
\small\normalsize
\begin{quote}
\caption{\label{fig:adding_bounces}
The average $\chi^2$ deviation of the $\phi_{i,N_{b}}$ from the uniform distribution for different frequency windows and different choices of $N_{b}$.  The $\chi^2$ were calculated for all possible frequency windows with a given width and then averaged over all different window realizations.  The random control was generated by a Monte-Carlo simulation of the $\chi^2$ for uniformly distributed phases.  Note that for the sixth bounce we exclude frequency windows in which the calculated $\boldsymbol{R}_{avg}$ falls below $0.1$.  (Color online)}
\end{quote}
\end{figure}

In Fig.~\ref{fig:adding_bounces}, we display the average $\chi^2$ statistic from our sample for multiple values of $N_{b}$ and different choices of window width.  The averaging is performed over all frequency windows of the same width, including windows whose frequencies overlap.  We see that for small window widths, increasing $N_{b}$ systematically decreases the error.  In addition, we also see that, up to a point, increasing the window size also decreases the error, but, once the error has leveled off, it decreases no further. This is consistent with the frequency averaging effectively removing the longer orbits, making the improved statistics from the larger $N_{b}$ irrelevant.  In addition, for comparison we include the $\chi^2$ statistics for a set of truly independent random phases and see that for the largest $N_{b}$ and widest window widths, our results are statistically indistinguishable from true randomness.

At this point we note one caveat to our use of Eq.~\eqref{eq:normed_theory_dist} relevant to Fig.~\ref{fig:adding_bounces}.  In Fig.~\ref{fig:fit_vs_bounces}, we see that $\boldsymbol{Z}_{t,6}$ drops almost to zero near 5.7GHz and 6.7GHz.  In fact, if we continue to add bounces, $\boldsymbol{R}_{t,N_{b}}$ can actually become negative at some frequencies, which would represent gain in a lossy system and is unphysical.  This unphysical behavior occurs only because the sum in Eq.~\eqref{eq:zeta_def} has been truncated, but it badly distorts the calculation in Eq.~\eqref{eq:normed_theory_dist} within the affected frequency ranges due to the abnormally small denominator.  This occurs because our sum over orbits is effectively an attempt to expand a function with poles near the real axis in a Fourier series.  Due to the rapidly changing features in $R_{avg,1,1}$, we get a form of Gibbs phenomenon, in which a Fourier series attempting to fit a discontinuous function systematically overshoots the fitted function.  For the purposes of producing Fig.~\ref{fig:adding_bounces}, we simply ignored frequency windows which contained frequencies such that $R_{t,6}(f)<0.1 R_{R}(f)$.  It may be possible to avoid this problem by using a smarter method to expand Eq.~\eqref{eq:Z_avg}.

\subsection{Two-port tests}\label{subsec:two_port_stats}

To test the results from the two-port configuration, we first test the statistical properties of the diagonal elements of the two-port impedance, considered separately.  From Eqs.~\eqref{eq:impedance_final} and \eqref{eq:littleV_mag}, and remembering that the distribution of $\boldsymbol{\xi}$ is basis-independent, we find that the diagonal elements of the multi-port impedance, considered independently, should be Lorentian random variables with width $R_{avg,n,n}$ and median $X_{avg,n,n}$.  Thus the diagonal elements of $\boldsymbol{Z}$ are susceptible to the same analysis used in the single-port case.  When we perform this statistical analysis on both diagonal elements of $\boldsymbol{Z}$ considered independently, we get results essentially identical to those shown for the single-port case.

\begin{figure}
  \begin{center}
  \includegraphics[width=3in]{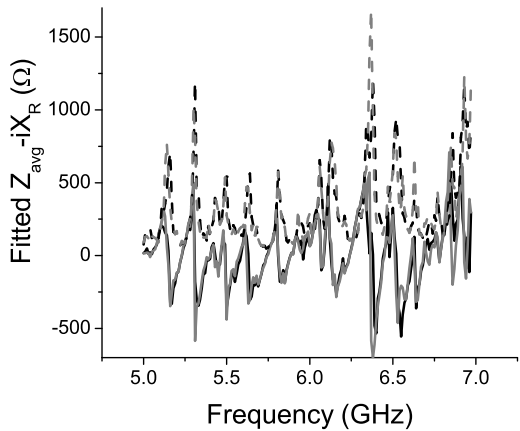}
  \end{center}
  \renewcommand{\baselinestretch}{1}
  \small\normalsize
  \caption{A comparison between the numerically found median (solid lines) and width (dashed lines) of the distribution of impedances at each frequency for the single-port simulated impedance (the black lines) and the (1,1) element of the two-port simulated impedance (the gray lines).  Note that for clarity we have subtracted the radiation impedance of the single port from the medians of both sets of data. }\label{fig:fitted_chi_values}
\end{figure}

Because port 1 is in the same location for both the single-port and two-port simulations, our theory also predicts that the median and width of the distribution of $Z_{1,1}$ at a single frequency should be almost identical to the median and width of the single-port impedance distribution at the same frequency.  The data is in agreement with this prediction as shown in Fig.~\ref{fig:fitted_chi_values} (note that, for the sake of clarity, we have subtracted the radiation reactance of the single port from the plotted medians of both sets of data).

The statistics of the two-port normalized impedance are more than the independent statistics of the diagonal elements; the elements of the $2\times 2$ matrix $\boldsymbol{Z}$ are strongly correlated.  All elements of $\boldsymbol{Z}$ go to infinity, for instance, when the frequency goes through a cavity resonance.  One common way of expressing this is via the correlation between the eigenvalues of $\boldsymbol{\xi}$.  The eigenvalues of $\boldsymbol{\xi}$ have the form $\cot(\theta_n/2)$, where the distribution of the $\theta_n$ is given by \cite{Brouwer_Lorentzian}
\begin{equation}\label{eq:eigenvalue_H_0_dist}
    P(\left\{\theta_n\right\})\propto \prod_{n<m}|e^{i\theta_n}-e^{i\theta_m}|^{\beta}.
\end{equation}
For the two-port impedance problem, this distribution simplifies to
\begin{equation}\label{eq:P_delta_phi}
    P(\theta_1,\theta_2)=\frac{1}{4}\left|\sin\left(\frac{\theta_1-\theta_2}{2}\right)\right|.
\end{equation}
Thus to test our theory, we must fit or calculate $\boldsymbol{Z}_{avg}$, find the values of $\boldsymbol{\xi}$ for our sample data, diagonalize, and find the distribution of the differences between the resulting phases.  We again use $\chi^2$ to determine the goodness-of-fit with the definition from Eq.~\eqref{eq:chi_2_def}, but with $\langle N_{r}\rangle$ determined by integrating Eq.~\eqref{eq:P_delta_phi}.  As we did for the single-port case, we both fit to find the numerical $\boldsymbol{Z}_{avg}$ and use the semiclassical sum.

Numerically fitting $\boldsymbol{Z}_{avg}$ is more difficult for the two-port case than in the one-port case.  We can find the diagonal elements of the fitted $\boldsymbol{Z}_{avg}$ simply by fitting the diagonal elements of $\boldsymbol{Z}$ to Lorenzians exactly as in the single-port case.  Fitting the off-diagonal elements numerically is more complex because both the shape and width of the distribution of the off-diagonal elements of $\boldsymbol{Z}$ depend in a non-linear way on all the elements of $\boldsymbol{R}_{avg}$.  Rather than attempting this more complex fit, we consider the rotated impedance matrix
\begin{eqnarray}\label{eq:rotated_impedance_final}
        \boldsymbol{O}\boldsymbol{Z}\boldsymbol{O}^{T}&=&i\boldsymbol{O}\boldsymbol{X}_{avg}\boldsymbol{O}^{T}\\&&
        +i(\boldsymbol{O}\sqrt{\boldsymbol{R}_{avg}}\boldsymbol{O}^{T})(\boldsymbol{O}\boldsymbol{\xi}\boldsymbol{O}^{T})(\boldsymbol{O}\sqrt{\boldsymbol{R}_{avg}}\boldsymbol{O}^{T}),\nonumber
\end{eqnarray}
where $\boldsymbol{O}$ is a constant orthogonal matrix.  Because the statistics of $\boldsymbol{\xi}$ are independent of basis, the diagonal elements of the rotated matrix $\boldsymbol{O}\boldsymbol{Z}\boldsymbol{O}^{T}$ will be Lorentzian distributed random variables with widths and means give by the diagonal elements of $\boldsymbol{O}\boldsymbol{X}_{avg}\boldsymbol{O}^{T}$ and $\boldsymbol{O}\boldsymbol{R}_{avg}\boldsymbol{O}^{T}$.  Thus if we make the simple choice
\begin{equation}\label{eq:O_def}
    \boldsymbol{O}=\begin{pmatrix}\frac{1}{\sqrt{2}} & \frac{1}{\sqrt{2}} \\ -\frac{1}{\sqrt{2}} & \frac{1}{\sqrt{2}} \\\end{pmatrix},
\end{equation}
we find that the widths of the distribution of the diagonal elements of the rotated impedance are
\begin{equation}\label{eq:ORO_diag}
    (\boldsymbol{O}\boldsymbol{R}_{avg}\boldsymbol{O}^{T})_{n,n}=\frac{R_{avg,1,1}+R_{avg,2,2}}{2}-(-)^{n}R_{avg,1,2},\\
\end{equation}
where $n=1,2$, and with corresponding logic for $\boldsymbol{X}_{avg}$.  Thus to find $Z_{avg,1,2}$, we fit the diagonal elements of $\boldsymbol{O}\boldsymbol{Z}\boldsymbol{O}^{T}$ to Lorenzians and take the half difference between the fitted medians and widths for the different diagonal terms.  This algorithm generalizes to larger numbers of ports by choosing $\boldsymbol{O}$ to rotate between the appropriate port indices.  In the process, we also confirm that the diagonal elements of the rotated impedance are in fact Lorentzian distributed, providing further support for Eq.~\eqref{eq:impedance_final}.

\begin{figure}
  \begin{center}
  \includegraphics[width=3in]{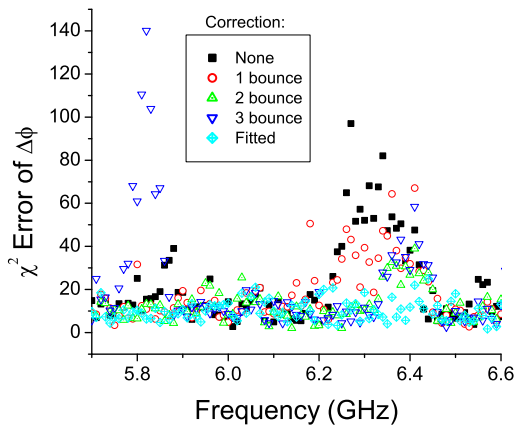}
  \end{center}
  \renewcommand{\baselinestretch}{1}
  \small\normalsize
  \caption{A comparison between the universality of the phases of $\boldsymbol{\xi}$ as extracted using different methods for normalizing the two-port impedance data.  All data shown is the $\chi^2$ statistics for fitting the different between phases to Eq.~\eqref{eq:P_delta_phi}. The gray lines represent error of the data normalized using the semiclassical impedance sum.  The solid line represents the phase difference statistics, but at each frequency using the fitted value for $\boldsymbol{Z}_{avg}$, where the fitting parameters are found as described in the text. (Color online)\label{fig:delta_phi_chi2}}
\end{figure}

In Fig.~\ref{fig:delta_phi_chi2}, we show the results of this analysis.  Rather than considering frequency windows, we simply consider the statistics of $|\theta_1-\theta_2|$ at each individual frequency using different values of $\boldsymbol{Z}_{avg}$.  We find that the numerically fitted $\boldsymbol{Z}_{avg}$ normalizes the data well, but not perfectly, with the range of $\chi^2$ values falling well within acceptable bounds as determined by the theoretical distribution of $\chi^2$, but systematically larger than would be expected from true randomness.  For the truncated sums, however, the story is rather different.  We see that at many frequencies, no corrections are needed at all to get good statistics.  Some frequencies, however, have large deviations from Eq.~\eqref{eq:P_delta_phi}.  That these deviations are caused by short orbits is demonstrated by the fact that, as we add 1 and 2 bounces to the semiclassical $\boldsymbol{Z}_{avg}$, the deviations initially become smaller, almost reaching the level of noise.  Unfortunately, adding longer orbits does not improve the situation; they leave the statistics either unchanged or markedly worse.  The reasons for this are unclear but are likely due to a combination of the Gibbs phenomenon observed in Sec.~\ref{subsec:one_port_stats} and the tendency of some impedance matrices to be poorly conditioned due to systematically large values of $\cot(\theta_n/2)$.

Thus we have confirmed many of the predictions of the extended random coupling model, including the approximate independence of the statistics of the diagonal elements of $\boldsymbol{Z}$, the invariance of the distribution under rotation, and the level spacing statistics for pairs of eigenvalues of $\boldsymbol{\xi}$.

\section{Adding Loss}\label{sec:adding_loss}

In practice, no real cavity will be truly lossless.  For non-zero frequencies, even cavities with superconducting boundaries have loss due to interactions between microwave photons and quasiparticles.  In quantum mechanical systems, there will always be dephasing, which is functionally equivalent to loss \cite{Brouwer_and_Beenakker}.  In this section we therefore address the effects loss has on our theory.

From Maxwell's equations, we derived in previous work \cite{Henry_paper_one_port} that going from lossless to uniformly lossy is performed by the transformation $k\rightarrow k+i\alpha$, where $\alpha=k/(2Q)$ and $Q\gg 1$ is the loss parameter of the closed cavity.  Performing this analytical continuation takes some care.  The function that must be explicitly continued analytically to obtain universal statistics of the sort we found previously is the normalized impedance $\boldsymbol{\xi}$, given by
\begin{equation}\label{eq:xi_def}
    i\boldsymbol{\xi}=\boldsymbol{R}_{avg}^{-1/2}\left(\boldsymbol{Z}-i\boldsymbol{X}_{avg}\right)\boldsymbol{R}_{avg}^{-1/2}.
\end{equation}
Because $\boldsymbol{R}_{avg}$ and $\boldsymbol{X}_{avg}$ appear independently in Eq.~\eqref{eq:xi_def}, we must analytically continue each independently.  Because taking the real and imaginary parts of non-constant functions is not an analytic operation, for lossy systems $\boldsymbol{R}_{avg}$ and $\boldsymbol{X}_{avg}$ are not real but rather the analytic continuation of the real and imaginary part of the lossless $\boldsymbol{Z}_{avg}$ on the real axis.  These analytic continuations are unique.

To find this analytic continuation for the microwave billiards used in our experiments, it is necessary to explicitly find the real and imaginary parts of $\boldsymbol{Z}_{avg}$.  For our microwave billiard system, we find that
\begin{eqnarray}\label{eq:R_explicit}
    \boldsymbol{R}_{avg}&=&\boldsymbol{R}_{R}+\boldsymbol{R}_{R}^{1/2}\boldsymbol{\rho}\boldsymbol{R}_{R}^{1/2} \\ \label{eq:X_explicit} \boldsymbol{X}_{avg}&=&\boldsymbol{X}_{R}+\boldsymbol{R}_{R}^{1/2}\boldsymbol{\chi}\boldsymbol{R}_{R}^{1/2}
\end{eqnarray}
where the $M\times M$ matrices $\boldsymbol{\rho}$ and $\boldsymbol{\chi}$ have the elements
\begin{eqnarray}\label{eq:rho_explicit}
    \rho_{n,m}&=&\sum_{b(n,m)}p_{b(n,m)}\sqrt{D_{b(n,m)}}\\&&\nonumber\times\cos\left(k\left(L_{p,n} + L_{p,m}+L_{b(n,m)}\right)-\pi/4\right),\\
    \chi_{n,m}&=&\sum_{b(n,m)}p_{b(n,m)}\sqrt{D_{b(n,m)}}\\&&\nonumber\times\sin\left(k\left(L_{p,n} + L_{p,m}+L_{b(n,m)}\right)-\pi/4\right),\label{eq:chi_explicit}
\end{eqnarray}
where we have made the empirically observed substitution $\Delta\phi_{n}=k L_{p,n}$, $L_{p,n}$ is observed to be a port-dependent constant, and $L_{b(n,m)}$ is the length of the trajectory $b(n,m)$, where we have used that for billiards, $S_{b(n,m)}(k)=k L_{b(n,m)}$.

Because $\boldsymbol{Z}_{R}$ and $D_{b(n,m)}$ change slowly in frequency compared to the level spacing,
they are approximately independent of $\alpha$ and thus equal to their lossless counterparts.  Thus the analytic continuations of $\boldsymbol{R}_{avg}$ and $\boldsymbol{X}_{avg}$ consist of keeping
the forms of Eqs.~\eqref{eq:R_explicit}-\eqref{eq:chi_explicit} unchanged but allowing $k$ to become complex.

After performing this continuation, $\boldsymbol{\xi}$ will no longer be real.  However the distributions of the real and imaginary parts of $i\boldsymbol{\xi}$ have been found as a function of the loss parameter, \cite{Lossy_approximate_impedance_Savin_Fyodorov,Lossy_impedance_Savin_Sommers_Fyodorov} and, for low-loss systems, we expect the distribution to be approximately universal.  The only difficulty with this analytic continuation is that the sum in Eq.~\eqref{eq:Z_avg_expansion} will not necessarily converge if the loss parameter is too high.  In such a case, it is necessary to perform the analytical continuation on the form of $\boldsymbol{Z}_{avg}$ given in Eq.~\eqref{eq:Z_avg}.  At this time, we have not attempted this, but we anticipate that it will require evaluating the denominator in Eq.~\eqref{eq:Z_avg} via a method other than short orbits, a poosible subject of further research.

We have experimentally tested this theory in the one-port case for a microwave quarter-bowtie billiard described in previous work for frequencies from 6 to 18 GHz \cite{Jen_hao_lossy_results}.  The primary difficulty in applying Eqs.~\eqref{eq:Z_avg_introduction} and \eqref{eq:zeta_def} over this frequency range is the fact that the loss parameter is not constant as a function of frequency.  However, our theory predicts that the phase of the normalized scattering parameter $s$ will be uniformly distributed independent of the loss parameter \cite{Henry_paper_one_port}, where $s$ is given by
\begin{equation}\label{eq:s_lossy_def}
    s=\frac{i\xi-1}{i\xi+1}.
\end{equation}
Thus by fitting the distribution of the phase of the normalized scattering parameter to the uniform distribution, we can again find the $\chi^2$ statistic for various frequency windows and different choices for the number of bounces before truncation $N_{b}$.  We have performed this experiment with a lossy cavity and found results qualitatively similar to Fig.~\ref{fig:adding_bounces} \cite{Jen_hao_lossy_results}.

\section{Conclusions}

In this paper, we have shown that the random coupling model, Ref.~\cite{Henry_paper_many_port} and Eq.~\eqref{eq:chaotic_impedance_statistics}, can be extended to take into account system specific short orbits that affect the statistical features of the system.  From the numerically and experimentally observed deviations of our results from universality, we anticipated that interactions between the walls and the port that were not sufficiently changed from realization to realization would result in corrections to our model.  We then derived a model that could predict such corrections.  Numerically and experimentally, we found that the improved model resulted in statistically significant improvement when fitting to the random coupling model, effectively reducing the deviations to the level of noise.  In addition, we developed utilizations of several mathematical tools, including Prange's semiclassical version of Bogomolny's $\boldsymbol{T}$ operator, that could be fruitful in further study of chaotic cavities and wave-chaotic systems with known dynamics in general.

\begin{acknowledgments}
This work was supported by the Air Force Office of Scientific Research Grant No. FA95500710049.  We greatly appreciate the thoughtful comments and input of Steven Anlage, Richard Prange and Schmuel Fishman.  In particular, we wish to dedicate this paper to the memory of Richard Prange (deceased), whose keen insights and generosity of spirit have been a constant inspiration.
\end{acknowledgments}

\end{document}